\documentclass{aa}
\usepackage{graphicx}
\usepackage{natbib}
\usepackage{txfonts}

\begin{document}

\title{Detection of seven 2+2 doubly eclipsing quadruple systems}

\author{Zasche,~P.~\inst{1},
        Henzl,~Z.~\inst{2,3},
        Ma\v{s}ek, M.~\inst{3,4},
        Uhla\v{r}~R.\inst{3},
        K\'ara, J.~\inst{1},
        Merc, J.~\inst{1},
        Ku\v{c}\'akov\'a,~H.~\inst{1,3,5,6}
        }

\offprints{Petr Zasche, \email{zasche@sirrah.troja.mff.cuni.cz}}

 \institute{
  $^{1}$ Charles University, Faculty of Mathematics and Physics, Astronomical Institute, V~Hole\v{s}ovi\v{c}k\'ach 2, CZ-180~00, Praha 8, Czech Republic\\
  $^{2}$ Hv\v{e}zd\'arna Jaroslava Trnky ve Slan\'em, Nosa\v{c}ick\'a 1713, Slan\'y 1, 274 01, Czech Republic \\
  $^{3}$ Variable Star and Exoplanet Section, Czech Astronomical Society, Fri\v{c}ova 298, 251 65 Ond\v{r}ejov, Czech Republic \\
  $^{4}$ FZU - Institute of Physics of the Czech Academy of Sciences, Na Slovance 1999/2, CZ-182 00, Praha, Czech Republic\\
  $^{5}$ Astronomical Institute, Academy of Sciences, Fri\v{c}ova 298, CZ-251 65, Ond\v{r}ejov, Czech Republic\\
  $^{6}$ Research Centre for Theoretical Physics and Astrophysics, Institute of Physics, Silesian University in Opava, Bezru\v{c}ovo n\'am. 13, CZ-746 01, Opava, Czech Republic\\
 }

\titlerunning{Detection of seven doubly eclipsing quadruples}
\authorrunning{Zasche et al.}

 \date{Received \today; accepted ???}

\abstract{In this work, we study a heterogeneous group of seven stellar systems for the first
time. Despite their different distances or spectral types, all of them belong to a very rare group
of quadruple systems of 2+2 architecture, where both of the inner pairs harbor eclipsing binaries.
These systems are: ASASSN-V J102911.57-522413.6 (inner periods 0.57272, and 3.79027 days), V1037
Her (0.78758 and 5.80348 days), WISE J181904.2+241243 (0.36713 and 0.41942 days), V2894 Cyg
(2.57434 and 1.30579 days), NSVS 5725040 (1.79368 and 0.76794 days), WISE J210230.8+610816
(1.84324 and 0.57159 days), and ZTF J220518.78+592642.1 (2.79572 and 3.34615 days). Their outer
mutual periods are: 9.3, 25.4, 18.7, 27.5, 2.6, 2.2, and 14.0 yr, respectively. These outer
periodicities were derived using longer time span of photometric observations of these systems and
analysing their period changes of both inner pairs via ETVs (eclipse-timing variations). Most of
these studied systems are detached, as evidenced by  the proper modelling of their light curves. A
few of them show significant eccentric orbits with apsidal motion (e.g. V2894 Cyg, and NSVS
5725040). Further spectroscopic follow-up observations would offer a better characterization of
the component star's parameters (for e.g. NSVS 5725040), as well as a potential interferometric
detection of the systems as real doubles on their mutual orbits (for e.g.  V1037 Her). A rather
interesting excess of systems close to a 3:2 mean motion resonance is seen only for early
spectral-type stars with higher temperatures.}

 \maketitle

\section{Introduction} \label{intro}

Many important findings derived from classical studies of eclipsing binaries (hereafter EBs) are
still relevant today. Despite the fact that these methods are about a century old, EBs still
represent useful tool for deriving many astrophysical parameters for stars and their orbits and
for studies of stellar populations, their formation mechanisms, stellar structures, evolution, and
so on (see e.g. \citealt{2012ocpd.conf...51S}, or \citealt{2021Univ....7..352T}).

The study of quadruples comprising of two eclipsing binaries with 2+2 architecture is still a
quite novel topic, since the first so-called doubly eclipsing system (V994~Her) was discovered by
\cite{2008MNRAS.389.1630L}. Having two distinct sources of eclipses, which can, in principle, be
modelled independently, there are many more constraints that should be taken into account. Thus,
independent analyses should lead to the same findings for distance, same age, same metallicity,
and so on. To prove its quadruple nature, we would have to confirm that both of them actually
orbit around a common barycenter via spectroscopy, interferometry, or eclipse-timing variation
(ETV) analysis signals of both inner eclipsing binaries. We chose the last method one for the
study presented here, given its long-term collection of photometric data that spans many years --
up to a few decades.

The group of doubly eclipsing systems showing two periods has expanded in recent years, now
counting more than 350 stellar systems. However, detailed analyses that have definitively proved
their architecture as a 2+2 quadruple are still relatively rare. These have mostly been  systems
on very short mutual orbits that show usually large dynamical interactions, published by group of
authors associated with T. Borkovits \& S. Rappaport
\citep{{2021MNRAS.503.3759B},{2018MNRAS.478.5135B},{2021ApJ...917...93K}}. These studies even
include the discovery of a sextuple system of three eclipsing binaries
\citep{2021AJ....161..162P}. In addition, there were have also been discoveries made by our group,
focusing mainly on systems with longer mutual orbital periods, carried out on the basis of
archival photometry and our own data \citep[see, e.g.][]{2019A&A...630A.128Z,
{2020A&A...642A..63Z}, {2022A&A...659A...8Z}}. The topic of close, dynamically interacting
multiples was comprehensively summarized in a recent review by \cite{2022Galax..10....9B}.

\section{The selected systems}

Our process for choosing these specific systems was  relatively straightforward. We tried to scan
many potential doubly eclipsing systems and attempted to identify the ones that obviously exhibit
some variations of period for both inner eclipsing pairs. Such variations in their eclipse times
have to be adequately covered for both A and B pairs, and have to be in opposition to each other
for A and B, respectively. This usually means that such a multiple system should also have both of
its eclipsing periods adequately observed in a range of older, ground-based data from different
databases to also be able detect the eclipses of both pairs  and to derive the eclipse times as
well. For some of the systems, this was quite problematic, especially as the data suffer from
large uncertainties.  We note that at least some indication of a movement of both pairs around a
common barycenter was detected. The necessity of all these systems being also visible in the
older, ground-based data led to slightly brighter systems (10-15mag), which are located in both
the southern and northern hemispheres.

The selected systems were chosen from our recent publication of new candidates of doubly eclipsing
stars showing two sets of eclipses \citep{2022A&A...664A..96Z}, along with one system from
publication by \citep{2021MNRAS.501.4669E}. Two others from our sample are presented here for the
first time as doubly eclipsing quadruples, namely, WISE J181904.2+241243, and NSVS 5725040. We
refer to Table \ref{systemsInfo} for the summary of basic information about these stars, their
various catalogue naming, and positions on the sky.

\begin{table*}
  \caption{Basic information about the systems.}  \label{systemsInfo}
  \scalebox{0.77}{
  \begin{tabular}{c c c c c c c}\\[-6mm]
\hline \hline
  Target name                 &  Other name            & TESS identification & RA [J2000.0]& DE [J2000.0]& Mag$_{max}$ $^{\star}$ &  Temperature information $^{\star\star}$       \\
 \hline
 \object{ASASSN-V J102911.57-522413.6} & 2MASS 10291156-5224135 & TIC 447369043      & 10 29 11.6 & -52 24 13.6 &  13.35 (V)     & T$_{eff} = 6268$ K   \\
 \object{V1037 Her}                    & TYC 2071-671-1         & TIC 286280289      & 16 56 56.9 & +29 19 06.5 &  11.91 (V)     & T$_{eff} = 5624$ K    \\
 \object{WISE J181904.2+241243}        & ATO J274.7676+24.2120  & TIC 84892676       & 18 19 04.2 & +24 12 43.3 &  13.60 (V)     & T$_{eff} = 5594$ K    \\
 \object{V2894 Cyg}                    & GSC 02682-00817        & TIC 104909909      & 20 02 25.5 & +35 40 11.5 &  10.60 (V)     & sp B5 \citep{1953ApJ...118...77A}  \\ 
 \object{NSVS 5725040}                 & TYC 3151-194-1         & TIC 11917056       & 20 17 25.7 & +39 17 36.6 &  11.79 (V)     & sp B1V \citep{2019ApJS..241...32L} \\ 
 \object{WISE J210230.8+610816}        & 2MASS 21023087+6108166 & TIC 305635022      & 21 02 30.9 & +61 08 16.7 &  14.73 (V)     & T$_{eff} = 9012$ K   \\
 \object{ZTF J220518.78+592642.1}      & 2MASS 22051878+5926420 & TIC 327885074      & 22 05 18.8 & +59 26 42.1 &  13.55 (V)     & T$_{eff} = 16695$ K    \\
 \hline
\end{tabular}}\\
 {\small Notes: $^\star$ - Out-of-eclipse magnitude, $V_{mag}$ taken from UCAC4 catalogue \citep{2013AJ....145...44Z}, or Guide Star Catalog II \citep{2008AJ....136..735L}.
 $^{\star\star}$ - Effective temperature taken from the Gaia DR3 catalogue \citep{2022arXiv220800211G} }
\end{table*}

\section{Photometric data used for the analysis}

The photometric data used in the current study can be divided into two parts. At first, these are
the super-precise data from Transiting Exoplanet Survey Satellite (TESS,
\citealt{2015JATIS...1a4003R}). These data were used for the light curve modelling of both inner
eclipsing binaries to derive their basic properties, such as relative radii, inclinations, and
fractional luminosities. These data were extracted from the TESS archive using the
{\tt{lightkurve}} tool \citep{2018ascl.soft12013L}. Typically, several TESS sectors of data are
available for each of the stars.

In addition, we also used older, ground-based archival photometry for certain stars. These data
are very useful for us when trying to trace the period variations of both pairs via the ETV
method. Much more scattered photometry than the TESS data provide us with a very useful source of
data thanks to their time spans, which  often go back several decades. Without these data, it
would be very difficult to prove the ETV and definitively confirm the mutual movement of both
binaries only using the TESS archive.

In addition, several dozens of nights of observation for these targets were also carried out for
the purposes of this study. These heterogeneous data were secured at several observatories: 1.
Ond\v{r}ejov observatory in Czech Republic, using a 65-cm telescope and G2-MII CCD camera equipped
with standard $V$ and $R$ photometric filters; 2. Danish 1.54-m telescope on La Silla in Chile,
remotely controlled, using the $R$ and $I$ filters; 3. FRAM 25-cm telescope located on La Palma
(Observatorio del Roquede de los Muchachos, see \citealt{2019ICRC...36..769P}); 4. FRAM 30-cm
telescope located in Argentina (part of the Pierre Auger observatory, see
\citealt{2021JInst..16P6027A}); 5. Three different private observatories in Czech Republic, using
smaller telescopes, with observers from the team of co-authors: M.Ma\v{s}ek, R.Uhla\v{r}, and
Z.Henzl.

For the reduction of all these data,  standard procedures using  dark frames and flat fields were
used, and the photometry was derived using standard aperture-photometry tools. All of these
photometric data points were only used for calculating precise times of eclipses for tracing the
ETVs with a higher degree of conclusiveness. All of these dedicated observations of the stars are
plotted as red symbols in the figures included throughout this paper.

\section{Analysis}

For the light-curve (hereafter LC) modelling, we used the well-known programme {\sc PHOEBE}
\citep{2005ApJ...628..426P}, which is originally based on the Wilson-Devinney algorithm
\citep{1971ApJ...166..605W}. However, having no radial velocities for our object, several
simplifying assumptions had to be made prior to fitting the individual LCs. For example, the issue
of the mass ratio and its derivation solely from the photometry can usually be problematic for the
detached binaries, as has been previously stated elsewhere in the literature (see e.g.
\citealt{2005Ap&SS.296..221T}). For this reason, we usually fixed its value to 1.0 for most of our
detached binary systems. In addition,  the synchronicity parameters were kept fixed at 1.0, while
the albedo and gravity brightening coefficients were also kept fixed at their suggested values,
according to the temperature.

The input temperature values for the primary components were taken from the latest Gaia DR3
catalogue \citep{2016A&A...595A...1G, 2022arXiv220800211G}, using the values from the GSP-Phot
pipeline. These values are summarised in Table 1.

We proceeded mostly step-by-step according to the following scheme. At first, using all
photometric data from the TESS satellite, we identified the more pronounced eclipsing pair (named
pair A), built its phased light curve, and attempted to carry out a preliminary fitting of this LC
shape. After subtracting this LC, we obtained a preliminary photometry for only pair B. Doing the
preliminary analysis also for this pair B, we  returned  to the complete photometry to re-analyse
the LC of pair A with the residual data. Afterwards, we returned to pair B again and made a better
fit to B. This iterative approach was repeated several times. When subtracting both A and B light
curves, the complete residuals should not show any evident phase-dependent variations. This is our
proof that the LC shapes are satisfactory.

However, as a second step, we needed to derive the individual times of eclipses (using our AFP
method, as described in \citealt{2014A&A...572A..71Z}). This method is using the light curves from
various databases or surveys, but phased with linear ephemerides over a longer time interval. This
has to be done due to the availability of only sparse photometry and combining the data over more
epochs makes the phased light curve adequately covered for the method. We usually used the time
interval of one year, but this can be arbitrarily changed with respect to the number of data
points in each interval. With such an approach, we derived a more suitable orbital period than the
one originally assumed. With the new period, the whole analysis and the LC modelling process
should be repeated again, and then iteratively several more times.

At first, we started with the simplest assumption, namely, that both A and B binaries contribute
the same to the total light curve (i.e. using a third light, $L_3 = 50\%$); thus, this parameter
could also be kept free for fitting. The sum of both luminosities for both pairs, $L_A + L_B$,
should give 1.00 (or 100\%) but, in reality, the total luminosity is usually over 1.0, simply due
to the additional light from nearby sources caused by large TESS pixels.

\section{Results}

In this section, we focus on the individual systems presented in our analysis in greater detail.

\subsection{ASASSN-V J102911.57-522413.6}

The first star in our study, ASASSN-V J102911.57-522413.6, is located in the constellation of
Velorum. This star was discovered as a variable one using the photometric data from the ASAS-SN
survey \citep{2014ApJ...788...48S,2017PASP..129j4502K}. However, these authors only detected the
shorter and more pronounced period of 0.573 days from their data. Additionally, the star was later
classified as doubly eclipsing in our recent study \citep{2022A&A...664A..96Z}, with another
periodicity of about 3.79 days found in the data. Both of the binaries show EA-type light curve,
indicating detached orbits. This is more evident for pair B, which is obviously eccentric having
the secondary eclipse located at the 0.58 phase from the primary one. No other detailed study of
the star has been published since then; also, its spectroscopic analysis is missing. The only
available Gaia spectrum \citep{2022arXiv220800211G} is  of poor quality, mainly due to its low
brightness (as a 13th magnitude star).

\begin{figure}
 \centering
 \begin{picture}(380,255)
 \put(0,120){
  \includegraphics[width=0.45\textwidth]{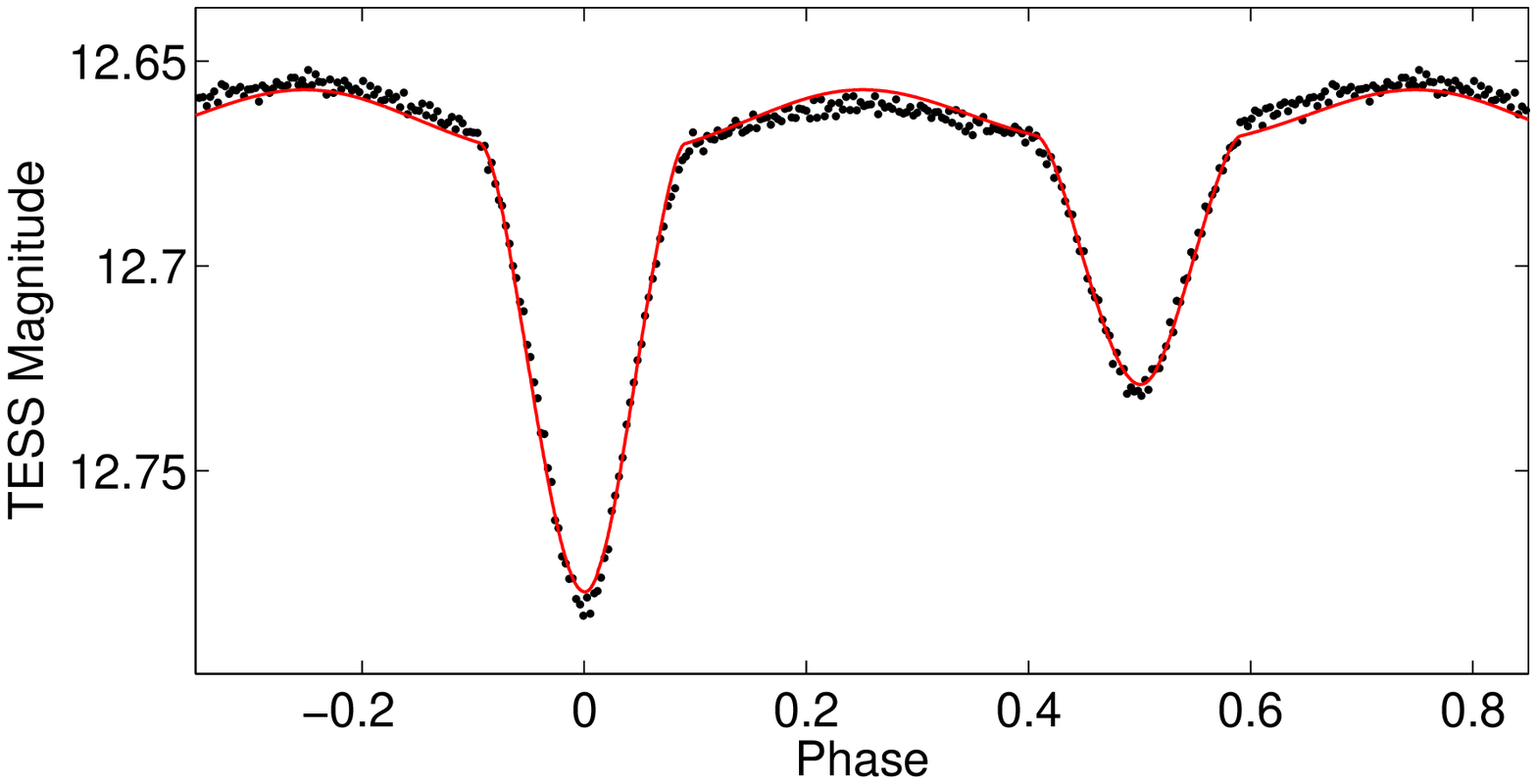}}
  \put(0,0){
  \includegraphics[width=0.45\textwidth]{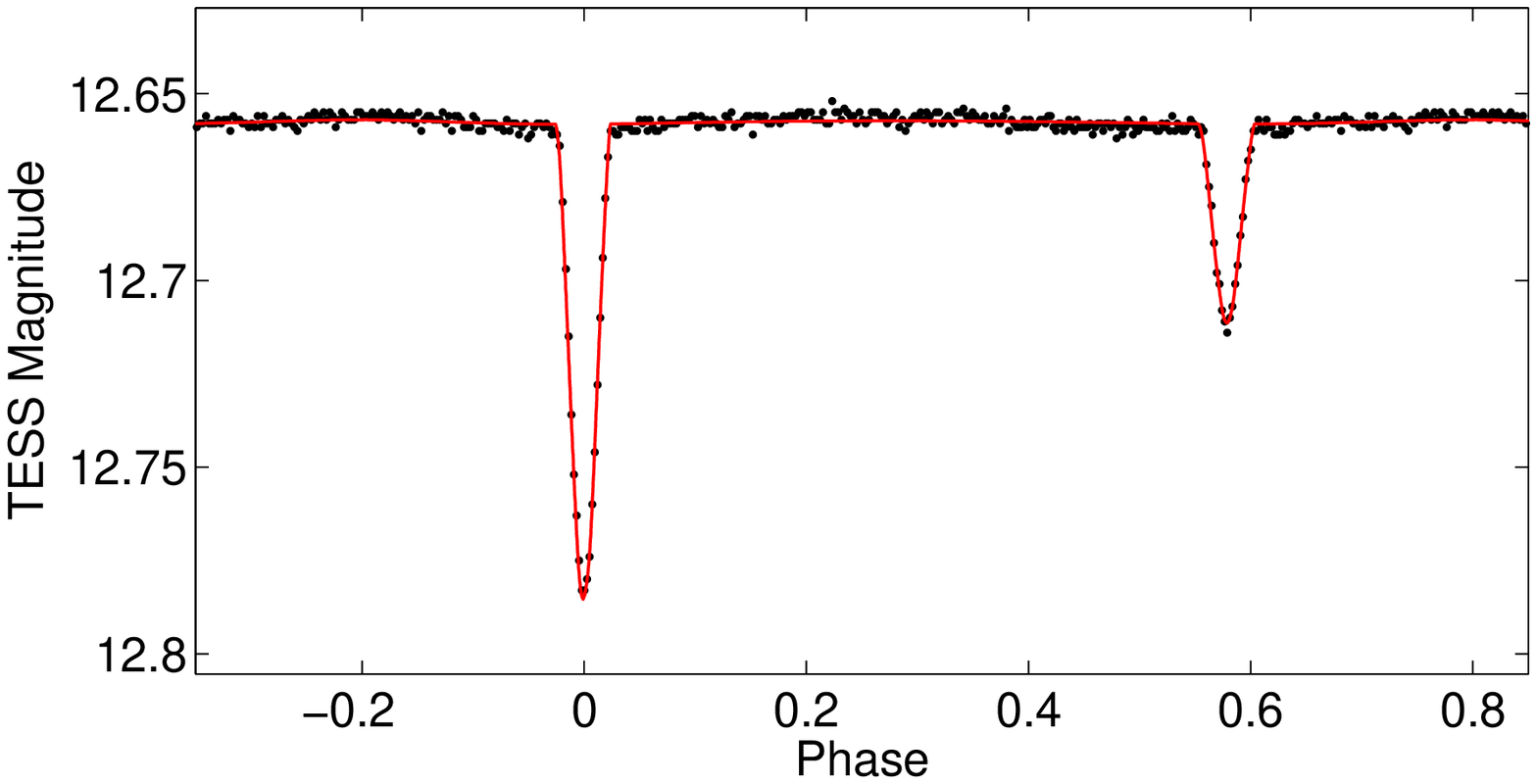}}
   \put(120,155){ {\hspace{11mm} \Large \textsf{Pair A}}}
   \put(120, 35){ {\hspace{11mm} \Large \textsf{Pair B}}}
  \end{picture}
  \caption{Light-curve fits of ASASSN-V J102911.57-522413.6 for both eclipsing pairs based on the PHOEBE fit and TESS data.}
  \label{FigLC_ASAS1029}
 \end{figure}

We chose the best available light curve for the system, namely, the TESS one from sector 37. These
data were analysed in {\sc PHOEBE}, resulting in the parameters given in Table \ref{TabLC}, and
the fits of the LC for both A and B pairs are given in Figure \ref{FigLC_ASAS1029}. As we can see,
there is an asymmetry of the LC of pair A, showing that near the quadratures of the orbit, the LC
has different brightness levels. Such a behavior is usually explained by a presence of surface
spots. We have not tried to fit them, we only wish to point out this peculiar aspect of the
system. Due to the significant out-of-eclipse variations,  the mass ratio was also fitted as free
parameter for pair A, while for the very detached system B, this was kept fixed during the fitting
process. We found quite a significant eccentricity for pair B, namely, about $e = 0.127$.

\begin{figure}
 \centering
 \includegraphics[width=0.45\textwidth]{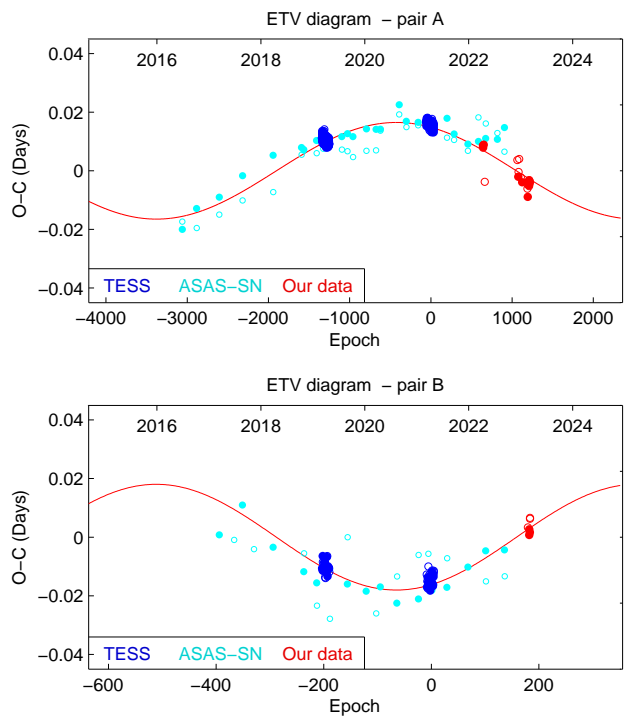}
 \caption{ETV diagram of ASASSN-V J102911.57-522413.6 showing two sets of eclipses of A and B pairs as resulting from our analysis.
 The theoretical curve representing the fit given in Table \ref{TabETV} is plotted as a red solid curve, while the dots stand for
 primary eclipses and the open circles for secondary ones. The color denotes the different source of data with the legend in the lower part of the plot.}
 \label{Fig_OC_ASAS1029}
\end{figure}

For studying the long-term evolution of the orbital periods for both pairs we collected the
available photometric data for the system spanning several years into the past (mainly the ASAS-SN
survey). However, some of the data do not show any photometric variation at all (e.g. ASAS or old
digitalized photographic plates via the DASCH project), due to their scatter and overly low
amplitude in the photometric variations of both pairs. These data were complemented with our new
dedicated observations of the system from two observing sites. At first, we observed the star on
La Silla using the Danish 1.54-m telescope equipped with a CCD camera and using standard $R$
filter. Secondly, on several other nights, the target was also observed using the 30-cm FRAM
telescope located in Argentina at the Pierre Auger Observatory. We used these data only for
deriving the times of eclipses for detecting the ETV in both pairs. The result of this fitting is
plotted in Figure \ref{Fig_OC_ASAS1029}. However, our current data are still too limited and cover
only part of the orbit. Hence, we decided to use only a more simple description of the orbit using
a zero eccentricity. Therefore, we needed to expand the time base of our data and/or obtain new
observations of a much better quality. Only then would it be possible to derive its correct
eccentricity as well. The fitting has led to period of about nine years. We also find that the
amplitude of ETV for pair B is slightly higher than that of pair A. This result is in good
agreement with our LC modelling, showing that the more dominant in luminosity is the pair A,
indicating  its higher mass as well.

Using the parallax of the system given by \cite{2022arXiv220800211G} of 0.711 mas (i.e. a distance
of about 1400~pc) the predicted angular separation of the two binaries should be about 10~mas.
Unfortunately, with such a small angular distance, we cannot hope to resolve the double -- even
with speckle interferometry technique, it is not possible due to the low brightness of the star.
Thus, only new upcoming observations in the next few years will be able to more definitively prove
our hypothesis and derive the eccentricity of the orbit as well.

\subsection{V1037~Her}

The second star included into our set is V1037~Her, which was discovered to be a variable star by
\cite{2000AJ....119.1901A}, but who reported an incorrect period for it. It was later found that
the dominant variation shows a periodicity of 0.78758 days, and showing a significant light curve
of a detached eclipsing type with deep eclipses of about 0.3 magnitudes. Over the last two
decades, the star was observed several times by amateur astronomers, who derived a few precise
times of eclipses of this binary. Quite surprisingly, nobody noticed that there is also an
additional variation with a period of about 5.8 days, also showing rather deep eclipse of more
than 0.1 magnitude (despite the fact that the eclipse is clearly visible on older photometric data
from various surveys as well; see more details below).

\begin{figure}
 \centering
 \begin{picture}(380,255)
 \put(0,120){
  \includegraphics[width=0.45\textwidth]{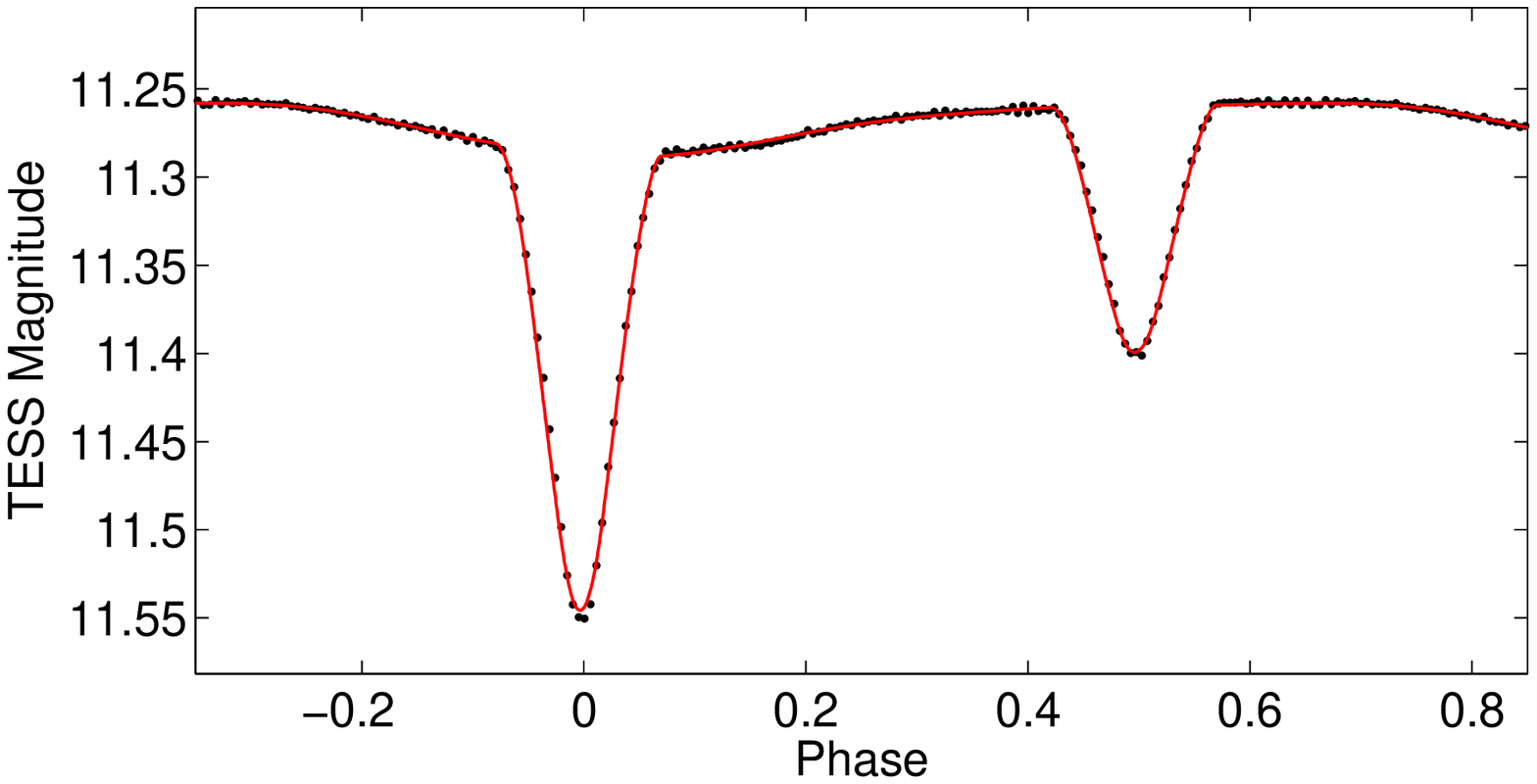}}
  \put(0,0){
  \includegraphics[width=0.45\textwidth]{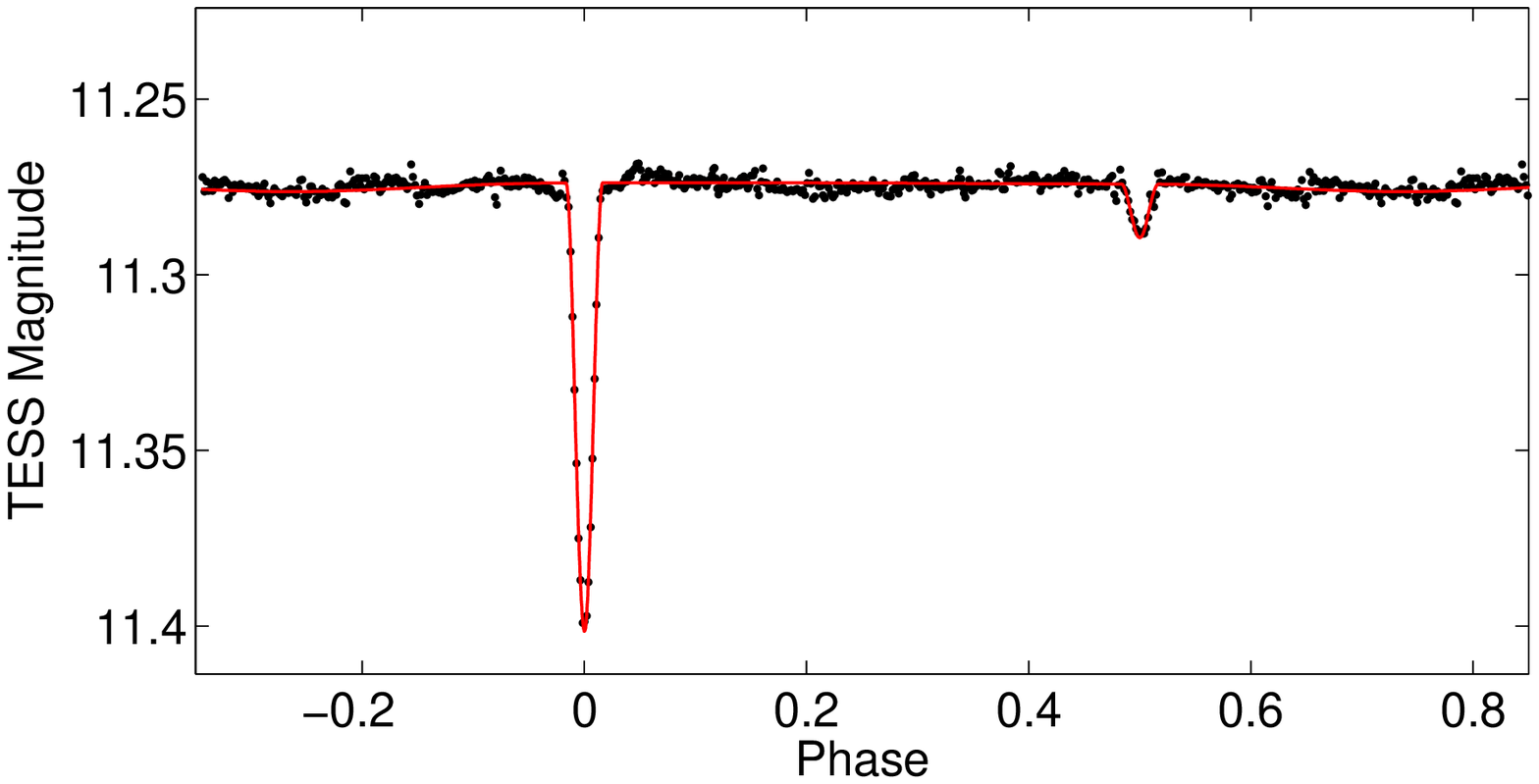}}
   \put(155,155){ {\Large \textsf{Pair A}}}
   \put(155,35){ {\Large \textsf{Pair B}}}
  \end{picture}
  \caption{Light-curve fits of V1037 Her for both A and B eclipsing pairs.}
  \label{FigLC_V1037Her}
 \end{figure}

As in the previous case, for the LC modelling, we used TESS data. In Fig. \ref{FigLC_V1037Her},
  our final fit of both LCs is shown, based on TESS satellite data. As we can see, the
secondary eclipses of pair B are only shallow, but clearly detectable in the TESS data. On the
other hand, pair A shows significant asymmetry, which is attributed to stellar spots. We used
the hypothesis of one spot located on the primary component, which is cooler than the surrounding
areas of the surface. Thus, we were  able to describe the asymmetry relatively well. Its parameters, as derived from
sector 52, are as follows:\ spot latitude -- 0.96~rad, longitude -- 5.04~rad, radius -- 0.16~rad, and temperature ratio --
0.66. However, when we compare these parameters with those of sector 25, we find that such a spot
cannot describe properly the LC and the asymmetry is different. Hence, an evolution of spot
parameters has to be taken into account when properly modelling the system. Both pairs are
circular, but  pair B has much more distant components, while  pair A has the components
much closer to each other, also showing a non-negligible ellipsoidal variation. Pair A is also
the dominant pair in the system concerning its luminosity level. The final LC parameters are given
in Table \ref{TabLC}.

\begin{figure}
 \centering
 \includegraphics[width=0.45\textwidth]{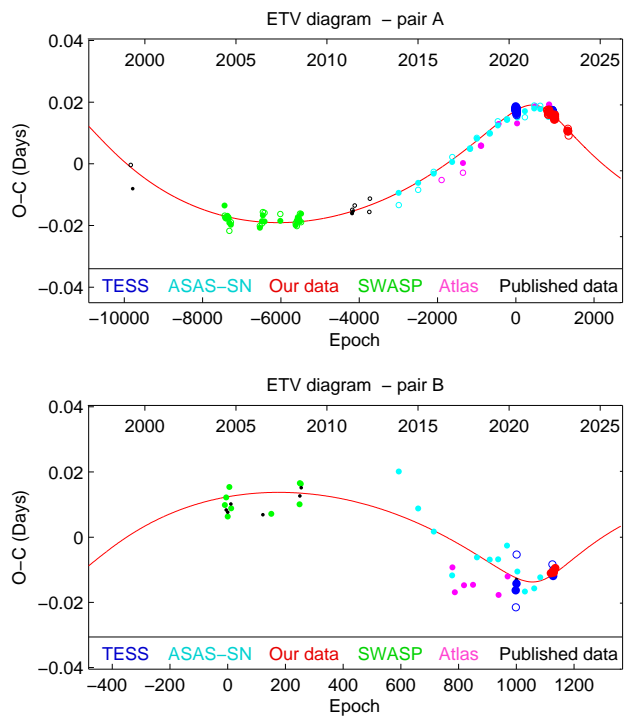}
 \caption{ETV diagram of both pairs of V1037 Her.}
 \label{Fig_OC_V1037Her}
\end{figure}

Besides the TESS photometry there were also many data points collected from various older
photometric databases and surveys, where both eclipsing periods are detectable as well. These are
mainly: the ASAS-SN survey, SuperWASP survey \citep{2006PASP..118.1407P}, and the ATLAS survey
\citep{2018AJ....156..241H}. Apart from these publicly available data, we also observed the star
over several nights with our own means. Our data were obtained in three different observatories:
the first is the private observatory of R.U. in J\'{\i}lov\'{e} u Prahy, CZ, using small 34-mm and
150-mm aperture telescopes and a standard $R$ filter; the second was carried out by M.M., using
the FRAM telescope of 25-cm diameter located in La Palma and using a standard $R$ filter;
finally, the third  is the private observatory of Z.H. in Velt\v{e}\v{z}e u Loun, CZ. All these
data were then used in Fig. \ref{Fig_OC_V1037Her}, where we can see the period variations of both
A and B pairs on their mutual orbit. Despite a quite longer period of about 26 years, which is
still not adequately covered by the data, we can confidently state that the system is bound and
both pairs orbit around each other. High eccentricity causes the rapid period variation that is
clearly visible near the periastron. From our orbital parameters and the distance to the system
from Gaia (d = 374~pc), we can compute the predicted angular separation of the two doubles on the
sky. This resulted in about 65 mas, which is much more favourable than for the previous case;
however, this is still at the edge of possible detection for such a star (i.e. of about 12
magnitudes).

\subsection{WISE~J181904.2+241243}

The next system studied here is WISE~J181904.2+241243, which was found to be an eclipsing binary
candidate by the ATLAS survey \citep{2018AJ....156..241H}. It shows a rather contact dominant pair
A, with a periodicity of about 0.367~days and about 0.3~mag deep eclipses; in addition, there is
also a weaker variation for pair B, which shows a slightly more detached configuration with period
of about 0.419~days. The star was not detected as doubly eclipsing before and this is the first
publication showing its true nature. WISE~J181904.2+241243 definitely exhibits the most contact
configuration (of both pairs) among the studied systems.

\begin{figure}
 \centering
 \begin{picture}(380,255)
 \put(0,120){
  \includegraphics[width=0.45\textwidth]{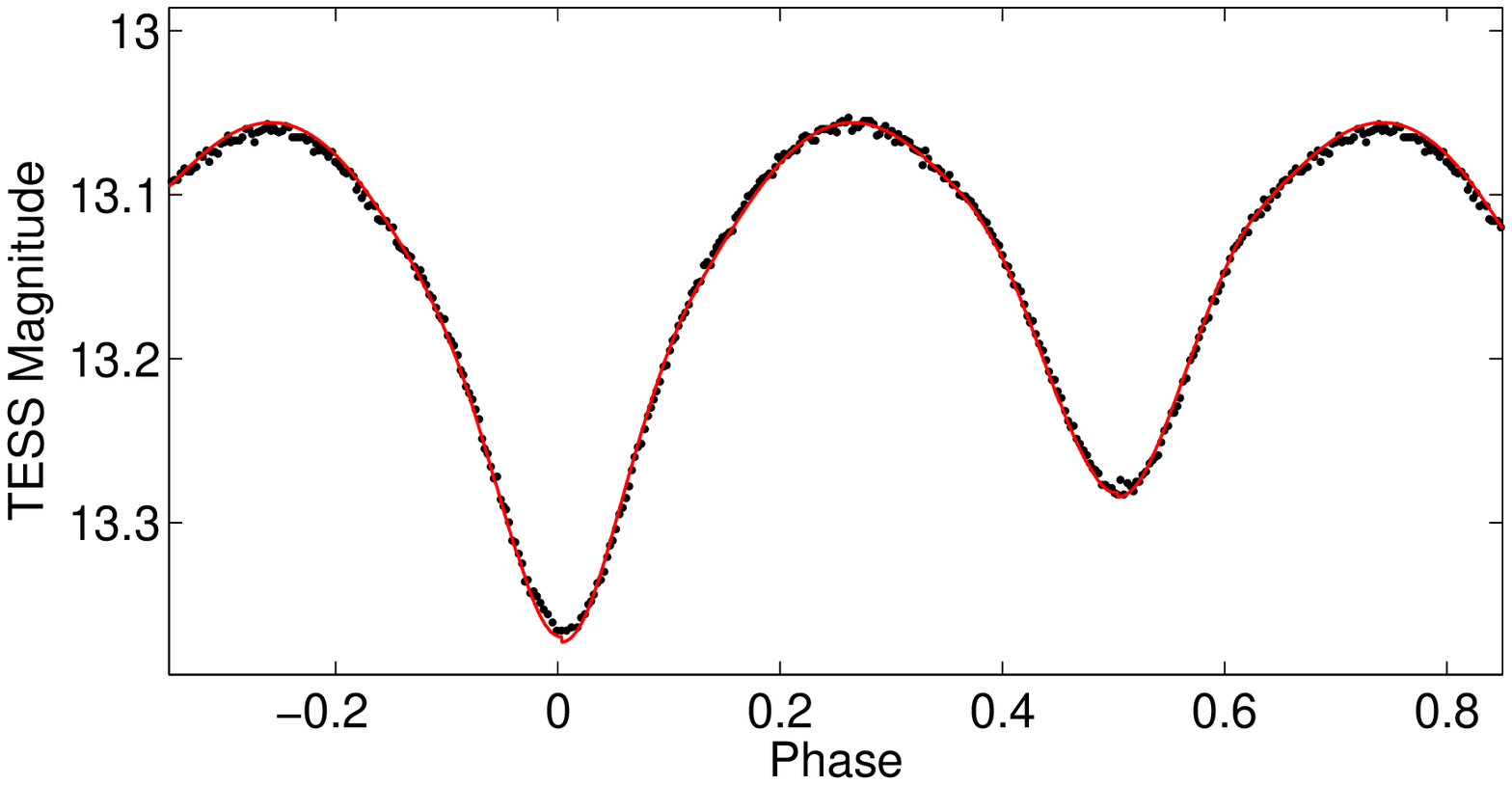}}
  \put(0,0){
  \includegraphics[width=0.45\textwidth]{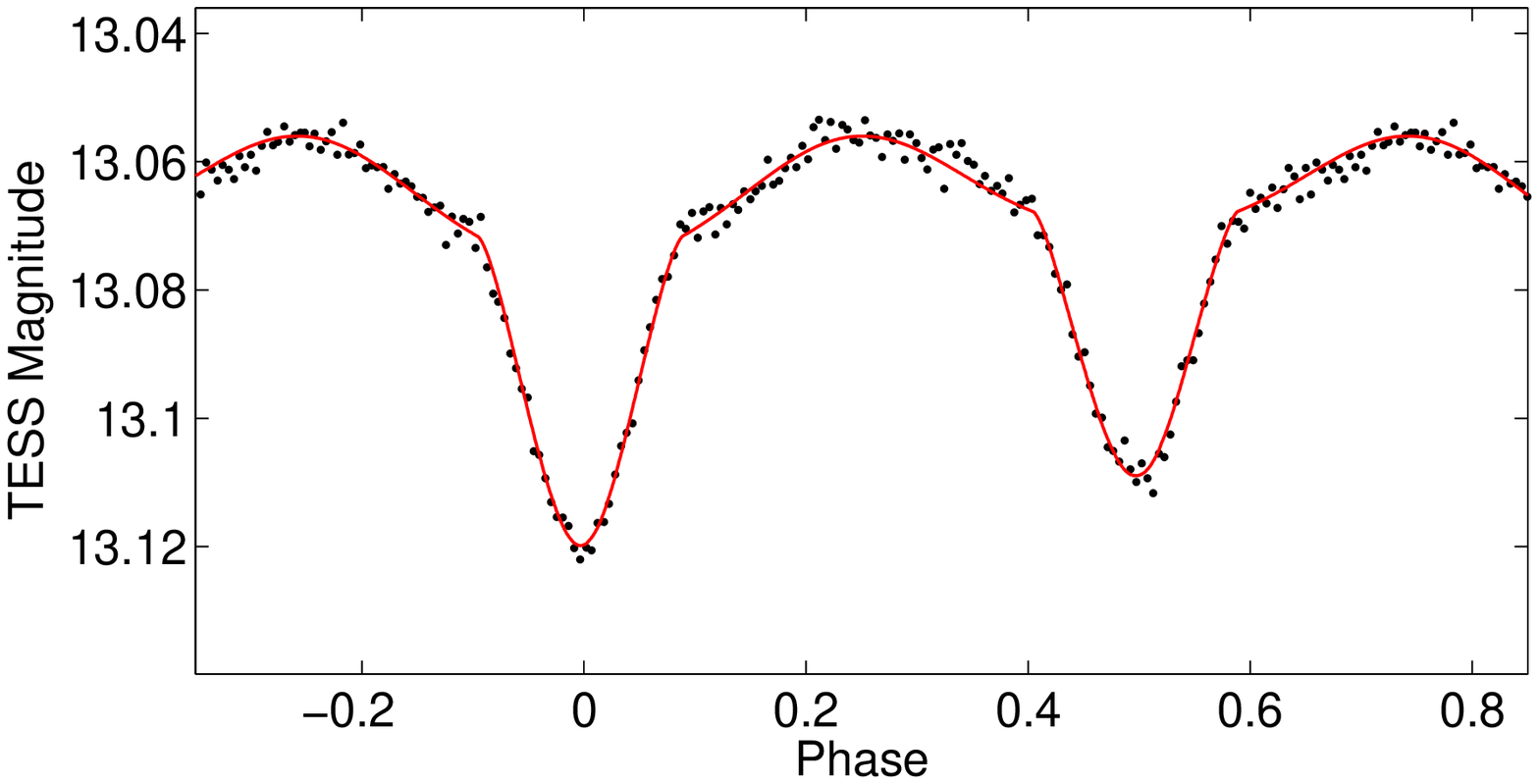}}
   \put(135,150){ {\Large \textsf{Pair A}}}
   \put(135,30){ {\Large \textsf{Pair B}}}
  \end{picture}
  \caption{Light-curve fits of WISE~J181904.2+241243 for both A and B eclipsing pairs.}
  \label{FigLC_WISE181904}
 \end{figure}

For the LC modelling, we used the TESS data from sector 40, where both eclipsing pairs are clearly
visible. The pair A resulted in a contact configuration of a W~UMa-type light curve. We also tried
to fit the mass ratio of this pair, since the ellipsoidal variations are large. The fraction of
luminosities indicates that the dominant is the pair A. However, pair B shows a slight
asymmetry of its light curve, which is moreover changing in between different sectors of data
making the whole analysis of its period changes more challenging. The results of the LC fitting
are given in Figs.\ref{FigLC_WISE181904}, while its parameters are given in Table \ref{TabLC}.

\begin{figure}
 \centering
 \includegraphics[width=0.45\textwidth]{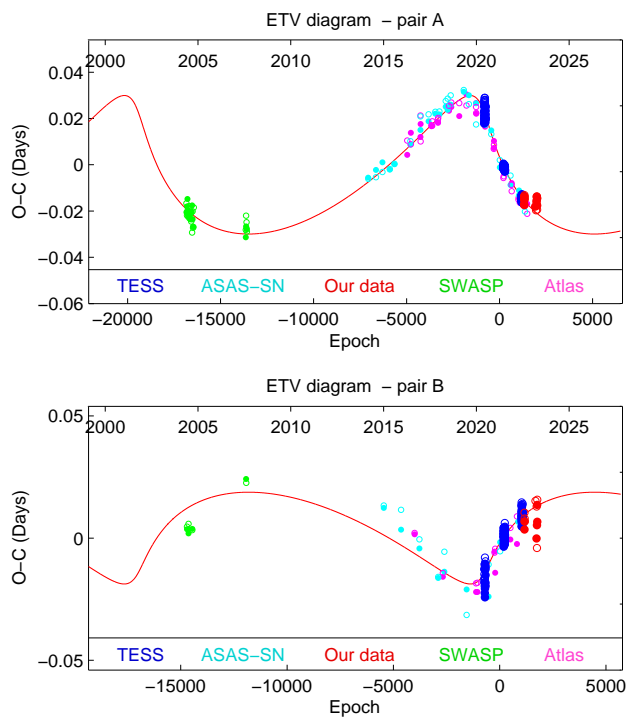}
 \caption{ETV diagram of both pairs of WISE~J181904.2+241243.}
 \label{Fig_OC_WISE181904}
\end{figure}

Concerning its period changes, we primarily used  the TESS photometry for deriving the eclipse
times of both pairs. In addition,  other databases such as ATLAS, ASAS-SN, and SuperWASP were
used. In particular, for pair A, these also provide us with quite precise estimates of times of
eclipses; however, for  pair B, due to its shallow eclipses, only more scattered datapoints were
derived. Figure \ref{Fig_OC_WISE181904} displays the result of the combined ETV fitting of both
pairs, while its parameters are given in Table \ref{TabETV}. As we can see, pair B resulted in
lower amplitude of ETV, indicating a higher mass than that of pair A. This is in contradiction to
the resulting luminosity ratios from the LC analysis. We have no clear explanation for such a
discrepancy. The most recent observations of both pairs should indicate some deviation from our
predicted light-time effect fit, so there is still a possibility that the overall orbit is
different than our presented solution. However, with the available data points, we were not able
to find a more suitable solution (even using the quadratic ephemerides).  Further observations in
the future should resolve this question.

 \subsection{V2894 Cyg}

Another system considered in our study is V2894 Cyg, which is the brightest star in our sample.
For this reason, it was also classified as a B5 star by \cite{1953ApJ...118...77A}. There was
sometimes a problematic identification of the star with a close-by star HD 227245, which has  a
similar brightness. However, we believe that the B5 spectral type belongs to our target. The star
was first detected as doubly eclipsing by \cite{2021MNRAS.501.4669E}, who gave its both eclipsing
periods of 1.306 and 2.575 days. The star is also probably a member of the open galactic cluster
named [FSR2007] 0198 \citep{2020A&A...633A..99C}.

\begin{figure}
 \centering
 \begin{picture}(380,255)
 \put(0,120){
  \includegraphics[width=0.45\textwidth]{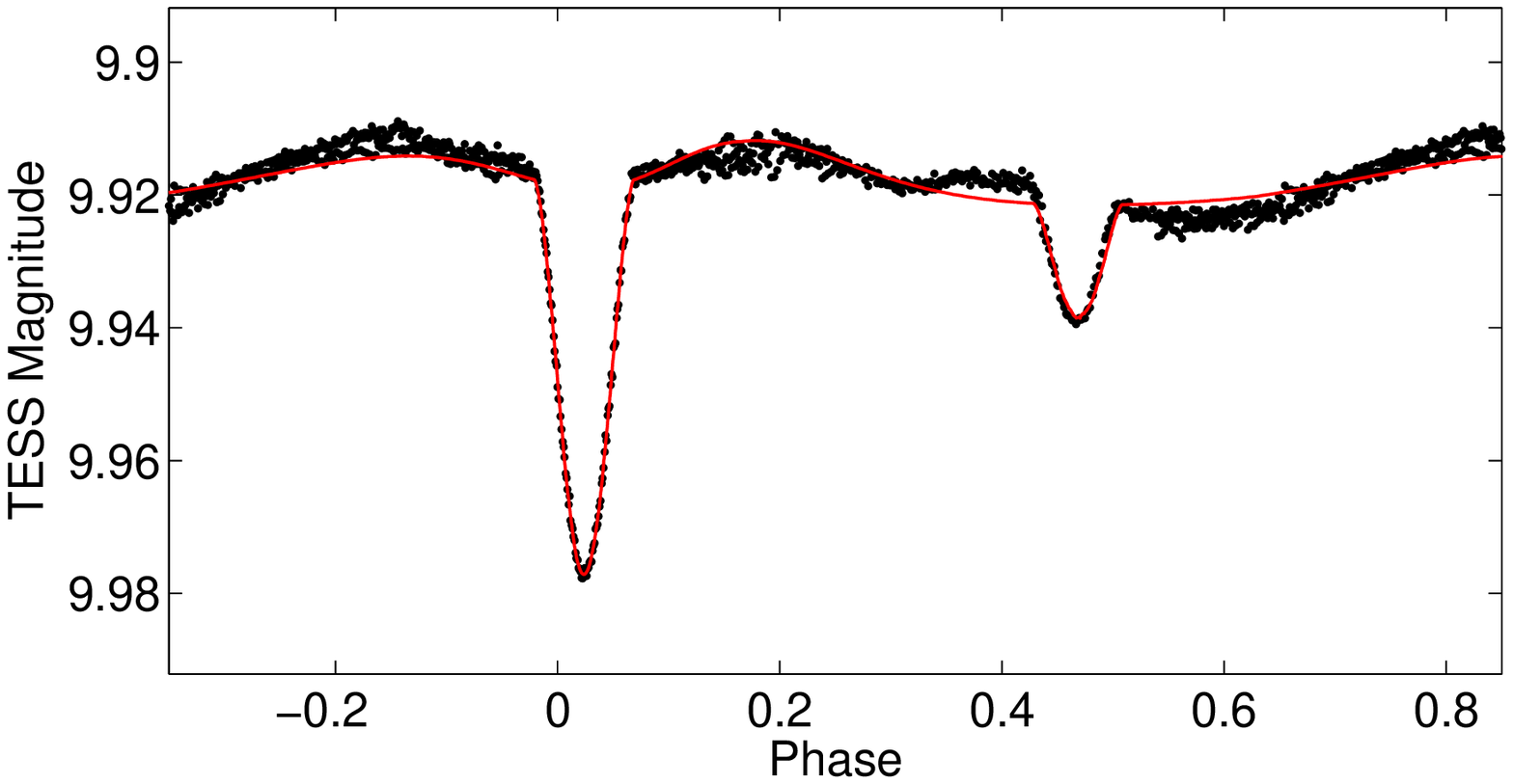}}
  \put(0,0){
  \includegraphics[width=0.45\textwidth]{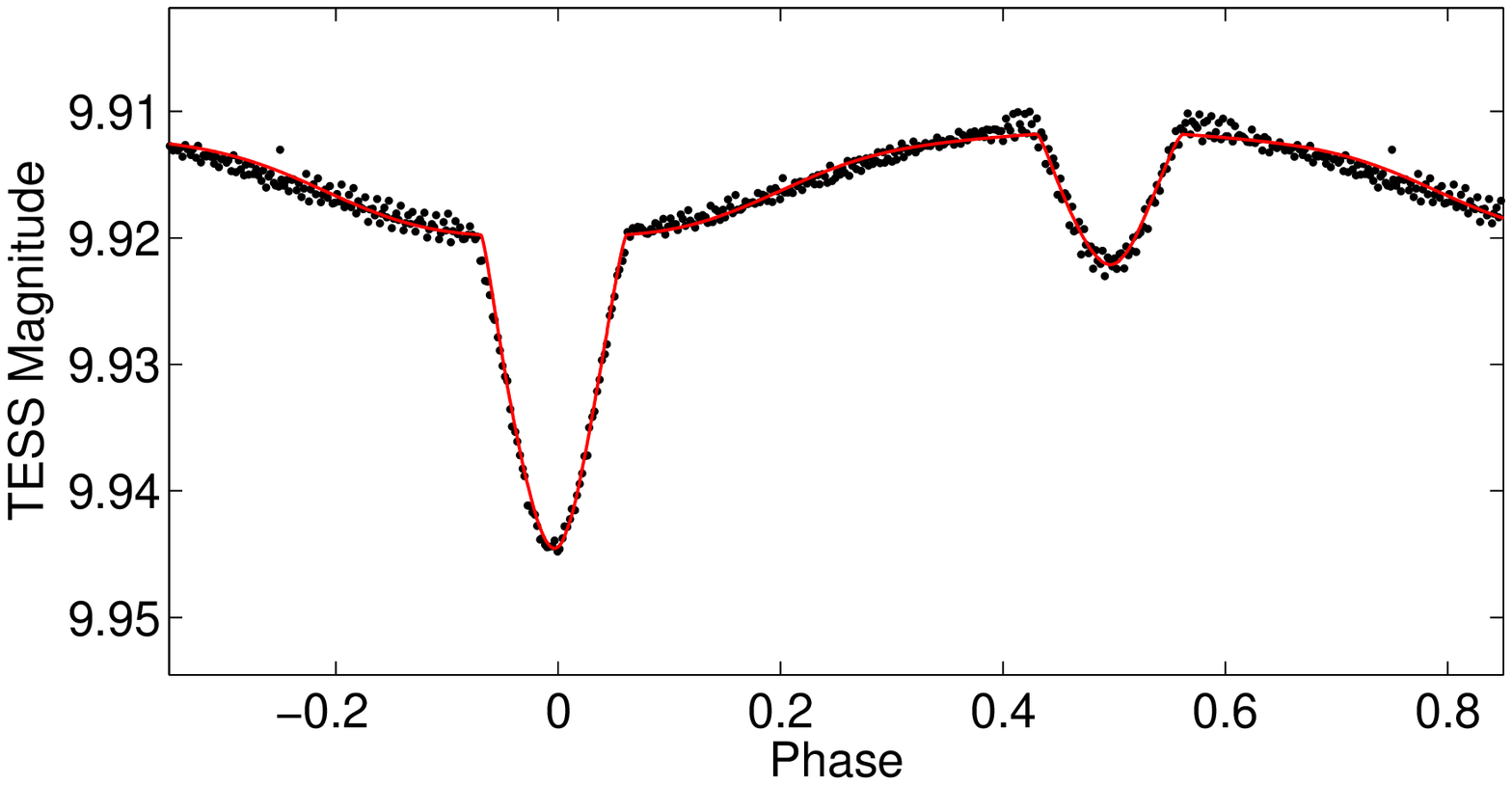}}
   \put(135,150){ {\Large \textsf{Pair A}}}
   \put(135,30){ {\Large \textsf{Pair B}}}
  \end{picture}
  \caption{Light-curve fits of V2894 Cyg for both A and B eclipsing pairs.}
  \label{FigLC_V2894Cyg}
 \end{figure}

We studied the star mainly using  the TESS data. The LCs of both pairs were analysed resulting in
the following picture. Pair A shows evident asymmetries, likely caused by some photospheric spots.
And moreover, the pair is also eccentric and showing significant apsidal motion. Such a movement
of apsides is visible even during the duration of the TESS data. Hence, this effect has to be
taken into account for a proper analysis. The results from our LC fitting are given in Table
\ref{TabLC}, while the final plots are shown in Fig. \ref{FigLC_V2894Cyg}.

\begin{figure}
 \centering
 \includegraphics[width=0.45\textwidth]{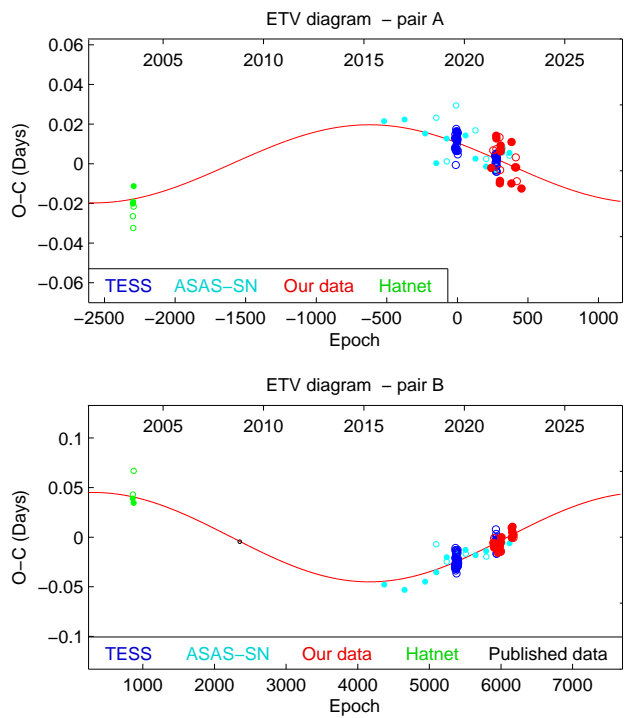}
 \caption{ETV diagram of both pairs of V2894 Cyg.}
 \label{Fig_OC_V2894Cyg}
\end{figure}

Due to an insufficient amount of data, we only applied a simplified approach of the circular outer
orbit, which is given in a plot in Fig. \ref{Fig_OC_V2894Cyg} and resulting parameters are given
in Table \ref{TabETV}. From the long-term variation of times of eclipses there resulted that the
apsidal motion has the period of about 46~years and the eccentricity of the orbit is 0.155. Such a
value is in great agreement with the 0.161 as resulted from the LC fitting. The plot shown in Fig.
\ref{Fig_OC_V2894Cyg} for pair A is plotted after subtraction of the long term apsidal motion,
only showing the contribution of the mutual orbit around a common barycenter. Mutual movement
shows that the orbit is relatively long, with a period of more than 27 years; hence, only part of
it is covered by the observations. New data in the coming years are therefore needed for a better
derivation of its orbital parameters.

\subsection{NSVS 5725040}

The next stellar system we studied is  NSVS 5725040, which has not previously been mentioned as a
doubly eclipsing system before; hence, it can be considered as a novel discovery. It is also the
second brightest star in our sample and due to its brightness, it also has a spectrum taken by the
LAMOST survey \citep{2015RAA....15.1095L}, which shows obviously the two-component feature. It was
also being classified as a B1V star based on \citep{2019ApJS..241...32L}. But no other more
detailed information about this system is available.

\begin{figure}
 \centering
 \begin{picture}(380,255)
 \put(0,120){
  \includegraphics[width=0.45\textwidth]{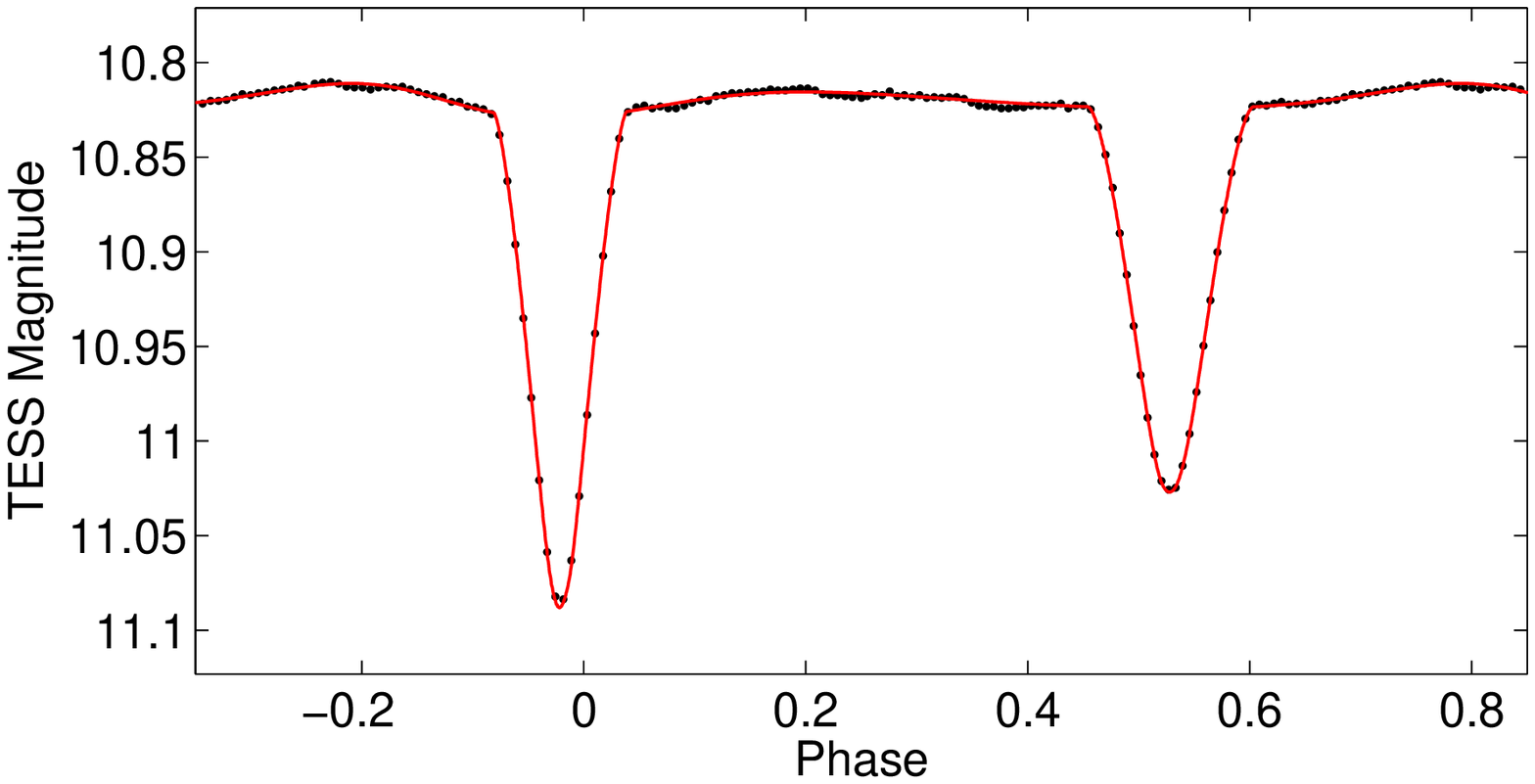}}
  \put(0,0){
  \includegraphics[width=0.45\textwidth]{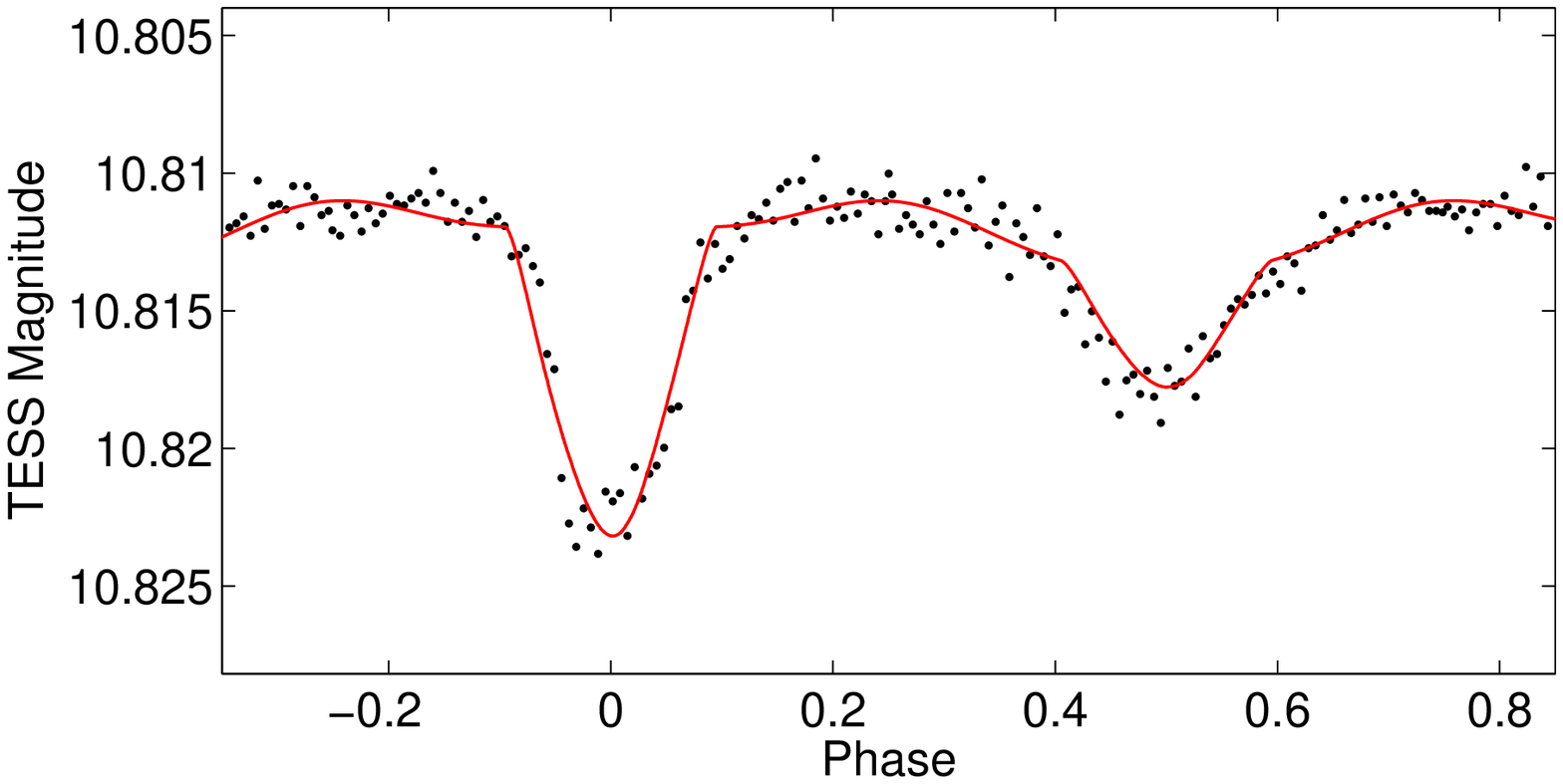}}
   \put(115,150){ {\Large \textsf{Pair A}}}
   \put(115,30){ {\Large \textsf{Pair B}}}
  \end{picture}
  \caption{Light-curve fits of NSVS 5725040 for both A and B eclipsing pairs.}
  \label{FigLC_NSVS5725040}
 \end{figure}

Using the TESS photometry we arrived at the following picture. At first, the more dominant pair A
shows an evidently eccentric orbit and also the fast apsidal motion. With an eccentricity of about
0.14 and orbital period of 1.79 days, it is among the systems with the highest eccentricity with
respect to stars with periods shorter than 2 days. The much shallower pair B has an orbital period
that is even shorter than pair A, but showing a circular orbit. All of the LC parameters are given
in Table \ref{TabLC}, while the shape of the LC fit can be seen in our Fig.
\ref{FigLC_NSVS5725040}. It is obvious that  pair A is very dominant in the luminosity levels,
which also causes pair B to only have such a low photometric amplitude of its eclipses, at the
level of only 0.01 mag.

\begin{figure}
 \centering
 \includegraphics[width=0.45\textwidth]{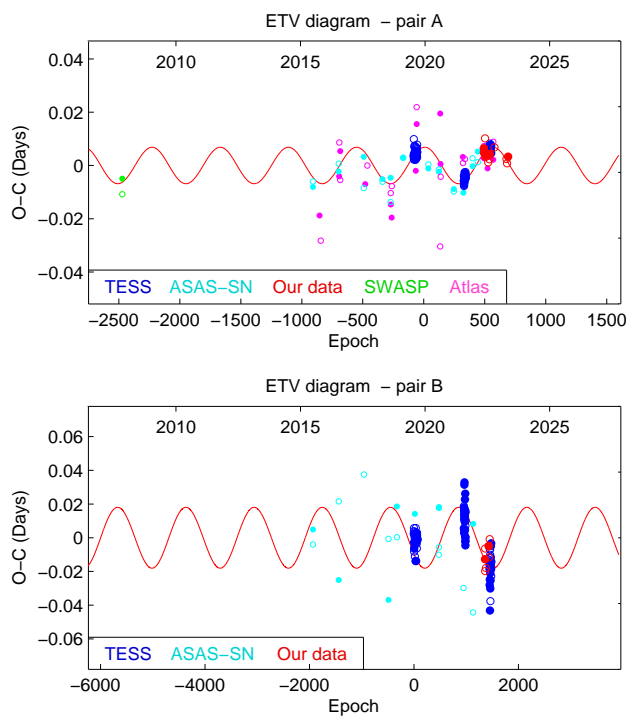}
 \caption{ETV diagram of both pairs of NSVS 5725040}
 \label{Fig_OC_NSVS5725040}
\end{figure}

Concerning the period changes and the ETV analysis, it was found quite problematic to analyze it
properly due to the very shallow eclipses of pair B, as this pair is almost undetectable in other
photometric data beyond TESS. There is also some older data going back to the beginning of the
20th century, but these are almost useless for our whole analysis. However, the period changes of
both pairs are pretty visible only using the TESS data due to the short period of the mutual A-B
orbit. The result of this analysis was that the eccentric orbit of pair A shows significant
apsidal motion with period of about 23.9 years only, which makes it one of the fastest apsidal
motion systems detected so far. Besides, the eccentricity of orbit A resulted in 0.140, in perfect
agreement with the eccentricity derived from the LC fitting (resulted in 0.141). The final fitting
of available data is shown in Fig. \ref{Fig_OC_NSVS5725040}. One remarkable consequence of our
result is the finding that the ratio of periods $p_{AB}^2/p_A$ resulted in about 1400 years. This
ratio shows us the level of these dynamical long-period interactions of the two inner and outer
orbits. For example, the period of nodal precession (if any, only in case of non-coplanar systems)
should be on the order of the same duration, that is, we should expect some variations of
inclination of pair A over a century of precise observations.

\subsection{WISE J210230.8+610816}

The star WISE J210230.8+610816 was detected as a doubly eclipsing system in our recent study
\citep{2022A&A...664A..96Z}. The dominant pair A has the period of about 1.84 days and the
amplitude of about 0.2 mag, showing obviously the light curve of detached configuration; while the
shallower pair B has the amplitude of about 0.06 mag only and with its period of about 0.57 days
shows a contact binary type light curve. It is the faintest star in our sample and, thus, no other
detailed study of the star has been published yet. Also, the spectrum of the star is not available
(besides the relatively poor Gaia spectrum).

\begin{figure}
 \centering
 \begin{picture}(380,255)
 \put(0,120){
  \includegraphics[width=0.45\textwidth]{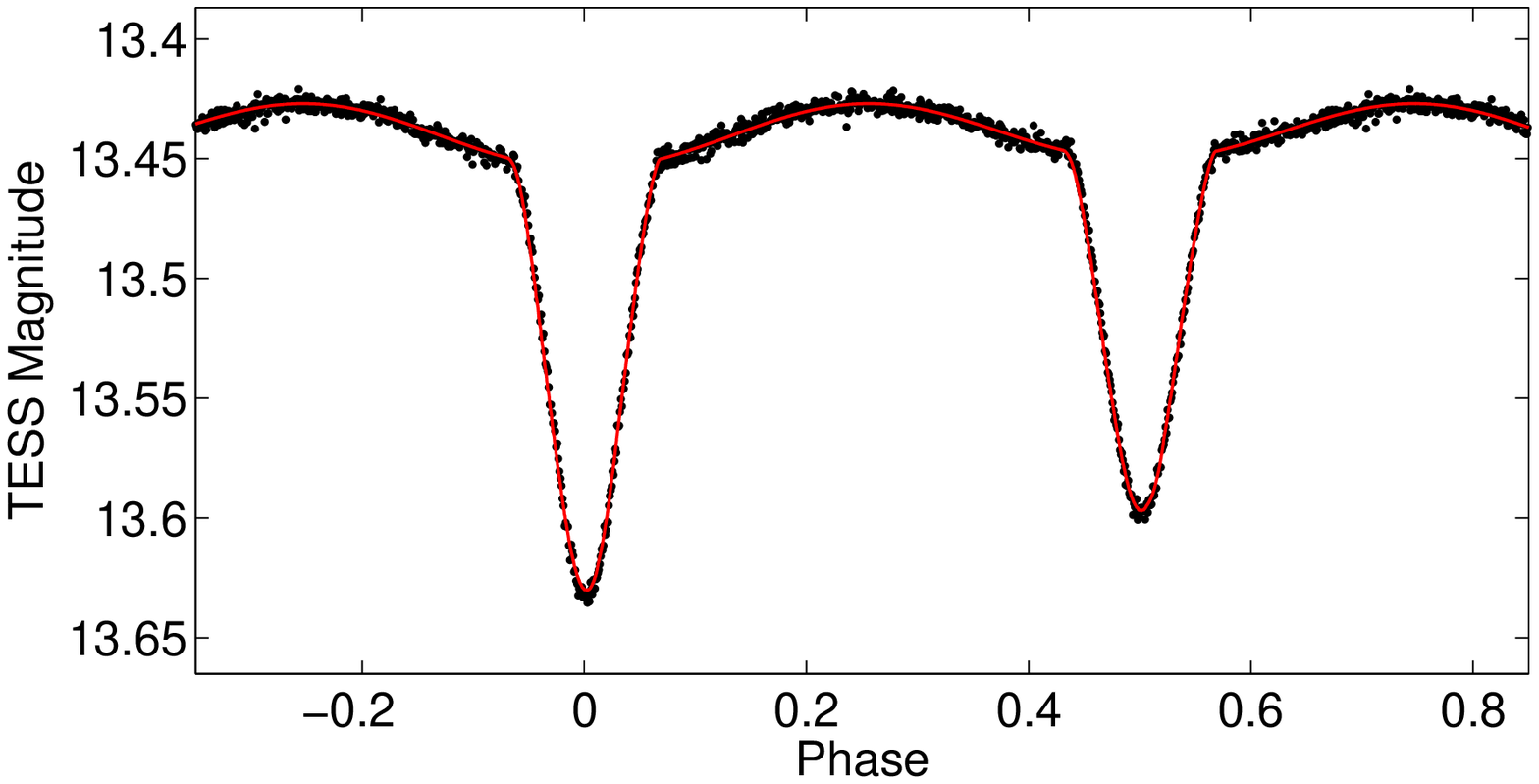}}
  \put(0,0){
  \includegraphics[width=0.45\textwidth]{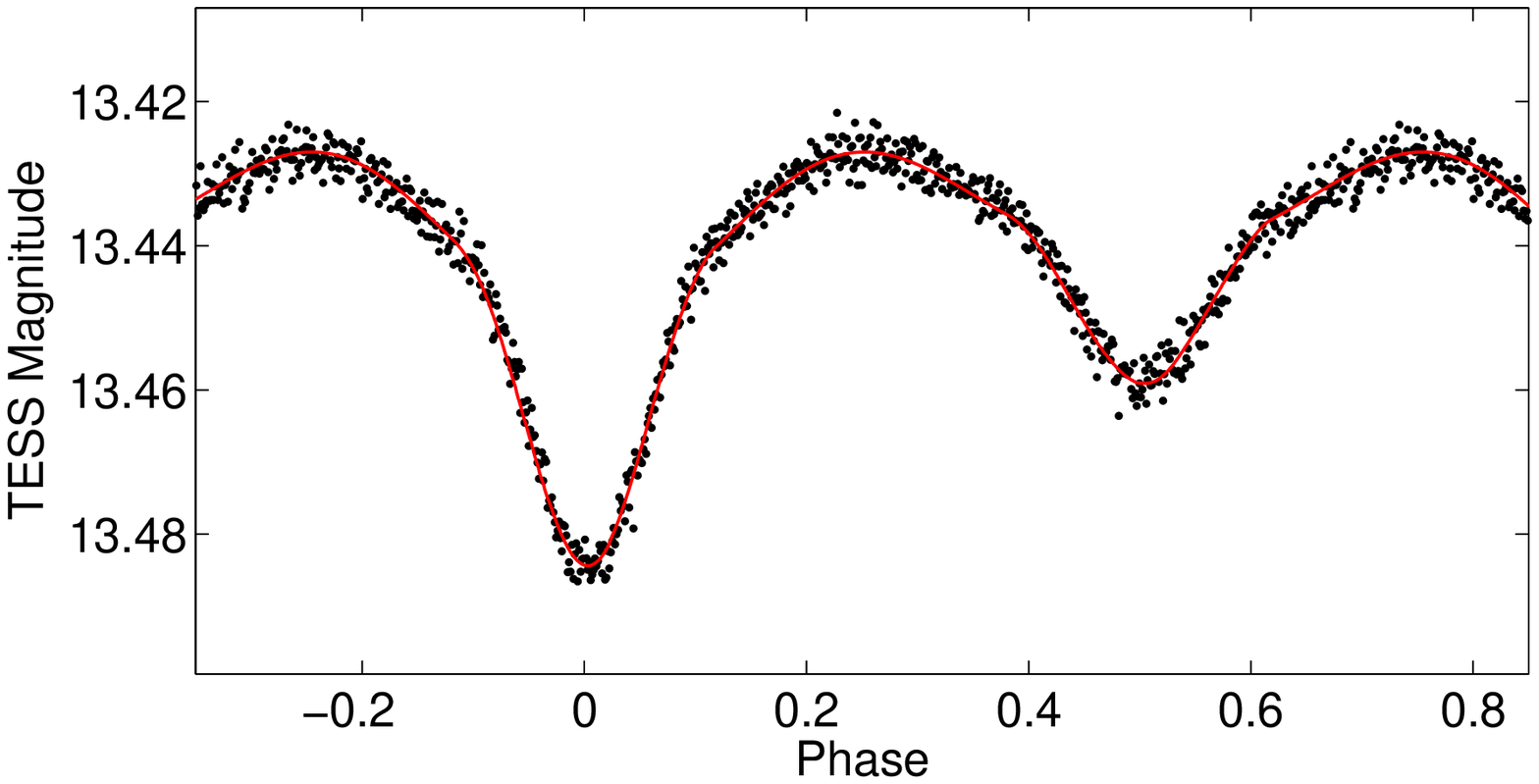}}
   \put(115,150){ {\Large \textsf{Pair A}}}
   \put(115,30){ {\Large \textsf{Pair B}}}
  \end{picture}
  \caption{Light-curve fits of WISE J210230.8+610816 for both A and B eclipsing pairs.}
  \label{FigLC_WISE210230}
 \end{figure}

We used mainly the TESS photometry to derive its basic physical and orbital properties. The star
was observed in six sectors and both sets of eclipses are clearly visible there. In our Fig.
\ref{FigLC_WISE210230}, we give  the final fits of both pairs as the result of  the PHOEBE
program. Its parameters are given in Table \ref{TabLC}, where one can clearly see that the pair A
is the dominant one. For this reason, the variations of pair B are quite problematic to detect in
other photometric databases and surveys. Pair B seems to be almost in contact, while pair A is a
detached one.

The study of period variation of both eclipsing pairs resulted in the following figure. The mutual
orbit is eccentric, having a period of only about 2.2 years, making it the fastest quadruple in
our sample. Amplitude of the pair B seems to be about 2 times larger, hence, also its mass should
be half that one of the pair A. Our result is in very good agreement with an independent finding
by \cite{2023A&A...670A..75C}, who analysed the Gaia and TESS data, giving the outer period value
of 843.88  $\pm$ 22.78 days. However, their study has not taken into account its quadruple nature,
since they did not detect the eclipses of the pair B in their data. Moreover, our finding is
supported with much larger dataset spanning a longer time interval than that  used in
\citeauthor{2023A&A...670A..75C}  Surprisingly,  the ratio of periods here, $p_{AB}^2/p_A$,
resulted in even lower value than for the previous system, namely, of only about 980 years. Hence,
we can hope detecting some inclination changes during the upcoming decades, in the case of
non-coplanar orbits.

\begin{figure}
 \centering
 \includegraphics[width=0.45\textwidth]{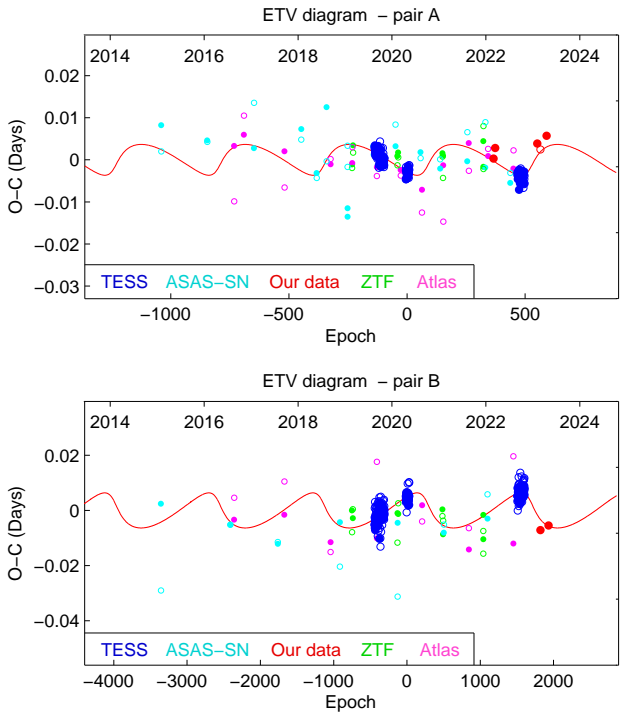}
 \caption{ETV diagram of both pairs of WISE J210230.8+610816}
 \label{Fig_OC_WISE210230}
\end{figure}

\subsection{ZTF J220518.78+592642.1}

The last system in our compilation is ZTF J220518.78+592642.1, which was first detected as a
doubly eclipsing system independently by \cite{2022A&A...664A..96Z} and
\cite{2022ApJS..259...66K}. The very dominant pair A has about a 2.8 d period and deep eclipses of
about 0.2 mag, while pair B is only of about 0.03 mag deep and having an approximately 3.3-d
orbital period. Besides the one Gaia spectrum, clearly showing a double-line profile, there are no
other spectra available for this star. With its inferred distance of about 7 kpc, this is the most
distant object among our stars.

\begin{figure}
 \centering
 \begin{picture}(380,255)
 \put(0,120){
  \includegraphics[width=0.45\textwidth]{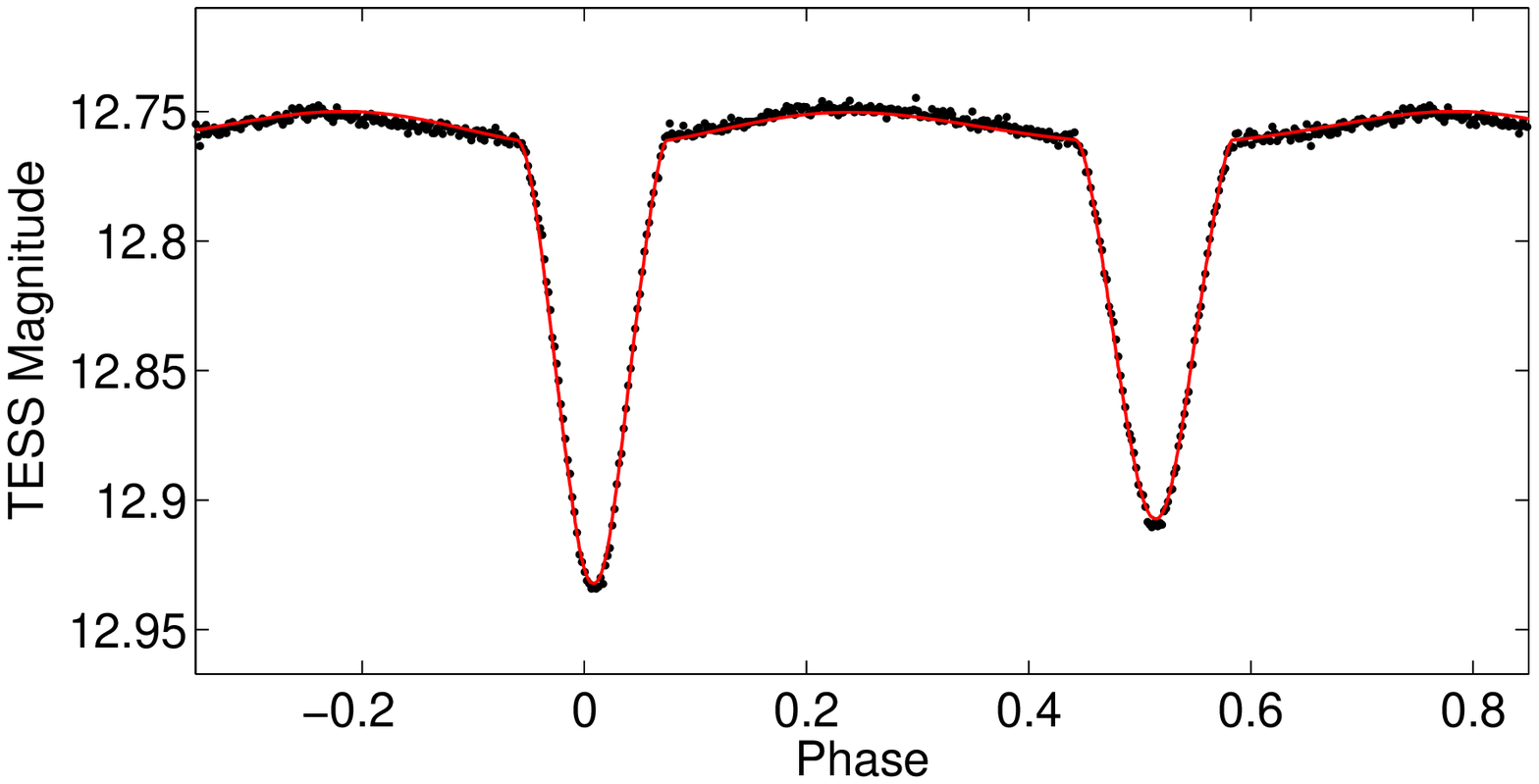}}
  \put(0,0){
  \includegraphics[width=0.45\textwidth]{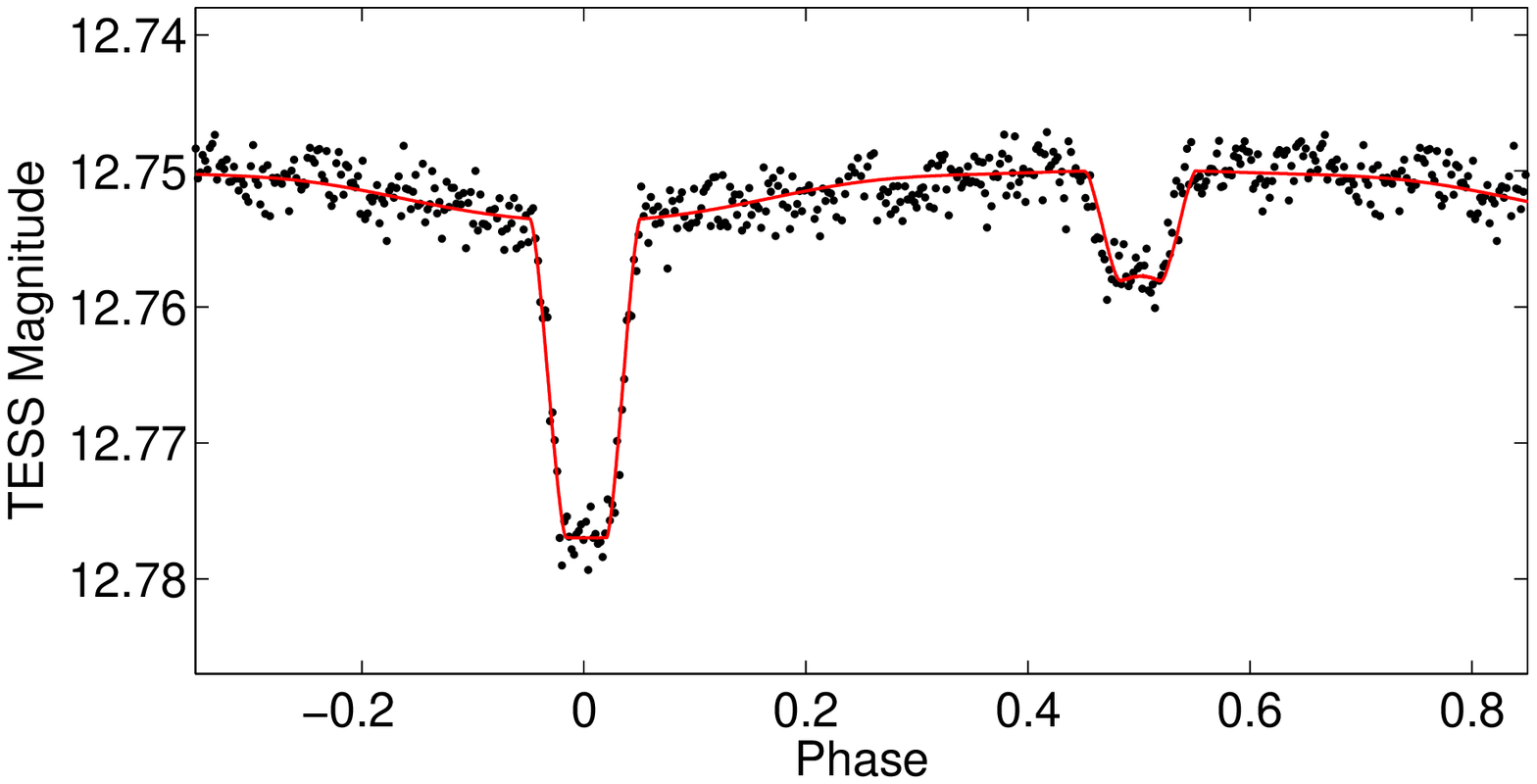}}
   \put(115,150){ {\Large \textsf{Pair A}}}
   \put(115,30){ {\Large \textsf{Pair B}}}
  \end{picture}
  \caption{Light-curve fits of ZTF J220518.78+592642.1 for both A and B eclipsing pairs.}
  \label{FigLC_ZTF220518}
 \end{figure}

Analysing its TESS light curves, we obtained the following results (also shown in Table
\ref{TabLC}, where are the parameters of the fit). Both the light curves are shown in Fig.
\ref{FigLC_ZTF220518}. As we can see, we deal with very dominating pair A (concerning its
luminosity), while pair B contributes only a few percent. Moreover, from the shape of the LC of
pair B, we see that it contains two rather different components (their temperatures). The pair A
also shows  a minimal asymmetry of its LC at quadratures.

\begin{figure}
 \centering
 \includegraphics[width=0.45\textwidth]{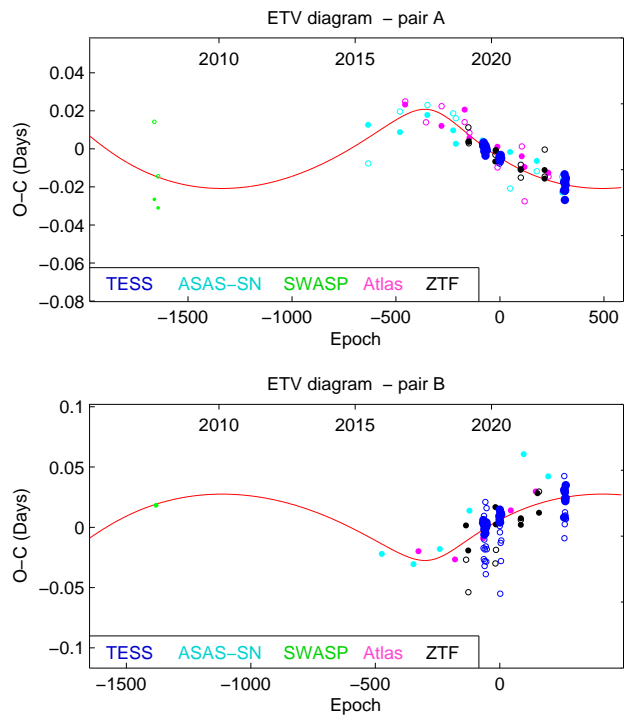}
 \caption{ETV diagram of both pairs of ZTF J220518.78+592642.1}
 \label{Fig_OC_ZTF220518}
\end{figure}

Collecting all available older measurements of the star, we carried out the long-term period
variation analysis. To detect  pair B is problematic due to its shallow eclipses. However, our
data clearly shows the period variations and both the ETVs behaving in the opposite manner. The
final parameters are given in Table \ref{TabETV}, while the fit is plotted in Fig.
\ref{Fig_OC_ZTF220518}. We find that pair B has higher amplitude for its ETV, but lower than we
would generally expect from the luminosity ratio as derived from the LC fitting. We leave this as
a open question, since the coverage of the ETV variation for pair B is still very poor and the
amplitude would also be much larger than our current fit shows.

\begin{table*}
\caption{Derived parameters for the two inner binaries A and B.}
 \label{TabLC}
 \tiny
  \centering \scalebox{0.82}{
\begin{tabular}{c | c | c | c | c | c | c | c}
   \hline\hline 
 \multicolumn{1}{c|}{System } & ASASSN-V J102911.57-522413.6 & V1037 Her  & WISE J181904.2+241243 &  V2894 Cyg         &   NSVS 5725040    & WISE J210230.8+610816 & ZTF J220518.78+592642.1 \\ \hline
  \multicolumn{1}{c|}{ }      &  \multicolumn{7}{c}{{\sc p a i r \,\, A}} \\ \hline
  $i$ [deg]                   &  69.48 $\pm$ 0.41     &  80.21 $\pm$ 0.26 &  69.98 $\pm$ 0.33     &  72.88 $\pm$ 0.35  & 84.70 $\pm$ 0.12  & 76.44  $\pm$ 0.09     & 76.76 $\pm$  0.24 \\
 $q=\frac{M_2}{M_1}$          &  1.02 $\pm$ 0.04      & 0.962 $\pm$ 0.027 &  0.95 $\pm$ 0.05      & 1.04 $\pm$ 0.03    &   1.0 (fixed)     &  1.00 (fixed)         &  1.0 (fixed)      \\
 $T_1$ [K]                    &   6268 (fixed)        &  5624 (fixed)     &  5594 (fixed)         &  12512 $\pm$ 134   &   26000 (fixed)   &  9012 (fixed)         &  16695 (fixed)    \\
 $T_2$ [K]                    &   5422 $\pm$ 142      & 4880 $\pm$ 119    &  4971 $\pm$ 84        &  15700 (fixed)     &  27580 $\pm$ 250  & 8525 $\pm$ 72         &  15477 $\pm$ 122  \\
 $R_1/a$                      &  0.315 $\pm$ 0.005    & 0.232 $\pm$ 0.005 &  0.391 $\pm$ 0.003    & 0.199 $\pm$ 0.004  & 0.244 $\pm$ 0.004 & 0.232 $\pm$ 0.005     &  0.236 $\pm$ 0.003 \\
 $R_2/a$                      &  0.302 $\pm$ 0.007    & 0.235 $\pm$ 0.005 &  0.381 $\pm$ 0.003    & 0.195 $\pm$ 0.004  & 0.176 $\pm$ 0.003 & 0.235 $\pm$ 0.004     &  0.230 $\pm$ 0.003 \\
 $L_1$ [\%]                   &  31.9 $\pm$ 0.6       &  38.5 $\pm$ 0.6   &  44.5 $\pm$ 1.1       & 21.6 $\pm$ 1.2     &  43.5 $\pm$ 2.1   & 39.1 $\pm$ 0.7        & 35.5 $\pm$ 0.9   \\
 $L_2$ [\%]                   &  14.4 $\pm$ 0.8       &  20.1 $\pm$ 0.5   &  26.4 $\pm$ 1.0       & 26.3 $\pm$ 1.4     &  17.9 $\pm$ 0.9   & 33.7 $\pm$ 0.6        & 32.9 $\pm$ 1.1   \\
 $L_3$ [\%]                   &  53.7 $\pm$ 1.2       &  41.4 $\pm$ 3.2   &  29.1 $\pm$ 1.8       & 52.0 $\pm$ 3.8     &  38.6 $\pm$ 2.4   & 27.2 $\pm$ 1.2        & 31.6 $\pm$ 2.3   \\ \hline
  \multicolumn{1}{c|}{ }      & \multicolumn{7}{c}{{\sc p a i r \,\, B}}\\ \hline
 $i$ [deg]                    &   85.69 $\pm$ 0.11    &  86.92 $\pm$ 0.24 &  60.84 $\pm$ 0.79     & 66.57 $\pm$ 0.18   & 63.47 $\pm$ 1.59  & 70.40 $\pm$ 1.33     & 81.54 $\pm$ 0.95 \\
 $q=\frac{M_2}{M_1}$          &  1.00 (fixed)         & 1.00 (fixed)      &   1.02 $\pm$ 0.04     & 0.98 $\pm$ 0.02    & 0.91 $\pm$ 0.23   &  0.734 $\pm$ 0.20    & 1.10 $\pm$ 0.08  \\
 $T_1$ [K]                    &   6268 (fixed)        & 5624 (fixed)      &  5594 (fixed)         &   15700 (fixed)    &   26000 (fixed)   &   9012 (fixed)       & 16695 (fixed)    \\
 $T_2$ [K]                    &   5535 $\pm$ 53       &  4003 $\pm$ 92    &  5219 $\pm$ 125       &  9396 $\pm$  140   &  25556 $\pm$ 160  &  6855 $\pm$ 157      & 8597 $\pm$ 198   \\
 $R_1/a$                      &  0.090 $\pm$ 0.002    & 0.071 $\pm$ 0.006 &   0.353 $\pm$ 0.010   & 0.264 $\pm$ 0.009  & 0.306 $\pm$ 0.011 & 0.379 $\pm$ 0.009    & 0.075 $\pm$ 0.006 \\
 $R_2/a$                      &  0.085 $\pm$ 0.002    & 0.057 $\pm$ 0.004 &   0.324 $\pm$ 0.011   & 0.284 $\pm$ 0.009  & 0.338 $\pm$ 0.006 & 0.318 $\pm$ 0.008    & 0.262 $\pm$ 0.010 \\
 $L_1$ [\%]                   &  24.1  $\pm$ 0.4      & 23.7 $\pm$ 1.3    &   26.7 $\pm$ 1.3      & 12.1 $\pm$ 0.9     &  4.3 $\pm$ 1.2    & 11.3 $\pm$ 0.9       &  2.1 $\pm$ 0.2  \\
 $L_2$ [\%]                   &   9.3  $\pm$ 0.4      &  1.9 $\pm$ 1.7    &   17.1 $\pm$ 1.0      &  4.7 $\pm$ 0.7     &  2.7 $\pm$ 0.7    &  3.7 $\pm$ 0.5       &  8.4 $\pm$ 0.5  \\
 $L_3$ [\%]                   &  66.6  $\pm$ 1.0      & 74.4 $\pm$ 2.6    &   56.2 $\pm$ 5.7      & 83.2 $\pm$ 1.9     & 93.0 $\pm$ 2.9    & 85.0 $\pm$ 1.3       & 89.5 $\pm$ 1.8  \\
 \noalign{\smallskip}\hline
\end{tabular}} \\
\end{table*}

\begin{table*}
\caption{Results of the combined ETV analysis.}
 \label{TabETV}
 \tiny
  \centering \scalebox{0.68}{
\begin{tabular}{c | c c c c c c c }
   \hline\hline 
 \multicolumn{1}{c|}{    }  & ASASSN-V J102911.57-522413.6 &       V1037 Her           &   WISE J181904.2+241243   &   V2894 Cyg                &   NSVS 5725040             &   WISE J210230.8+610816    & ZTF J220518.78+592642.1 \\ \hline
  $JD_{0,A}$ [HJD-2450000]  &  9312.5030 $\pm$ 0.0009      & 8992.6168 $\pm$ 0.0012    & 9312.6113 $\pm$ 0.0009    & 8725.5357 $\pm$ 0.0055     & 8831.2567 $\pm$ 0.0026     &  8965.0498 $\pm$ 0.0003    & 8966.5166 $\pm$ 0.0021  \\
 $P_A$ [day]                &  0.5727181 $\pm$ 0.0000009   & 0.7875781 $\pm$ 0.0000004 & 0.3671301 $\pm$ 0.0000002 & 2.5743422 $\pm$ 0.00000021 & 1.7936849 $\pm$ 0.0000110  &  1.8432444 $\pm$ 0.0000008 & 2.7957226 $\pm$ 0.0000064 \\
  $JD_{0,B}$ [HJD-2450000]  &  9315.1539 $\pm$ 0.00201     & 3196.5528 $\pm$ 0.0045    & 9312.6044 $\pm$ 0.0007    & 1685.4354 $\pm$ 0.0029     & 8686.9116 $\pm$ 0.0015     &  8964.1816 $\pm$ 0.0004    & 8964.3735 $\pm$ 0.0034 \\
 $P_B$ [day]                &  3.7902680 $\pm$ 0.0000123   & 5.8034722 $\pm$ 0.0000072 & 0.4194242 $\pm$ 0.0000002 & 1.3057949 $\pm$ 0.0000027  & 0.7679373 $\pm$ 0.0000018  &  0.5715872 $\pm$ 0.0000004 & 3.3461464 $\pm$ 0.0000101 \\
 $p_{AB}$ [yr]              &   9.26 $\pm$ 3.8             &      26.2 $\pm$ 5.3       &  18.7 $\pm$ 0.4           & 27.5 $\pm$ 4.1             &  2.64 $\pm$ 0.12           &  2.22 $\pm$ 0.07           & 14.01 $\pm$ 2.23 \\
 $A_A$ [d]                  &  0.0164 $\pm$ 0.0022         &     0.018 $\pm$ 0.002     &   0.030 $\pm$ 0.002       & 0.020 $\pm$ 0.005          & 0.007 $\pm$ 0.003          &  0.0037 $\pm$ 0.0009       & 0.021 $\pm$ 0.005 \\
 $A_B$ [d]                  &  0.0180 $\pm$ 0.0054         &     0.014 $\pm$ 0.008     &   0.019 $\pm$ 0.003       & 0.045 $\pm$ 0.014          & 0.018 $\pm$ 0.006          &  0.0064 $\pm$ 0.0007       & 0.028 $\pm$ 0.009 \\
 $e$                        &  0.0 (fixed)                 &     0.550 $\pm$ 0.009     &   0.697 $\pm$ 0.004       & 0.0 (fixed     )           &  0.0 (fixed)               &  0.542 $\pm$ 0.102         & 0.454 $\pm$ 0.010 \\
 $\omega$ [deg]             &  --                          &     104.9 $\pm$ 9.3       &   138.9 $\pm$ 1.8         & --                         &  --                        &  334.7 $\pm$ 24.8          & 104.7 $\pm$ 11.2 \\
 $T_0$ [HJD-2450000]        &  --                          &    9457.3 $\pm$ 1778.0    &   9003.1 $\pm$ 52.9       & --                         &  --                        &  9118.1 $\pm$ 44.0         & 6890.5 $\pm$ 452.0 \\
 \noalign{\smallskip}\hline
\end{tabular}} \\
\end{table*}

 \begin{figure*}
 \centering
 \begin{picture}(510,250)
 \put(0,120){\includegraphics[width=0.48\textwidth]{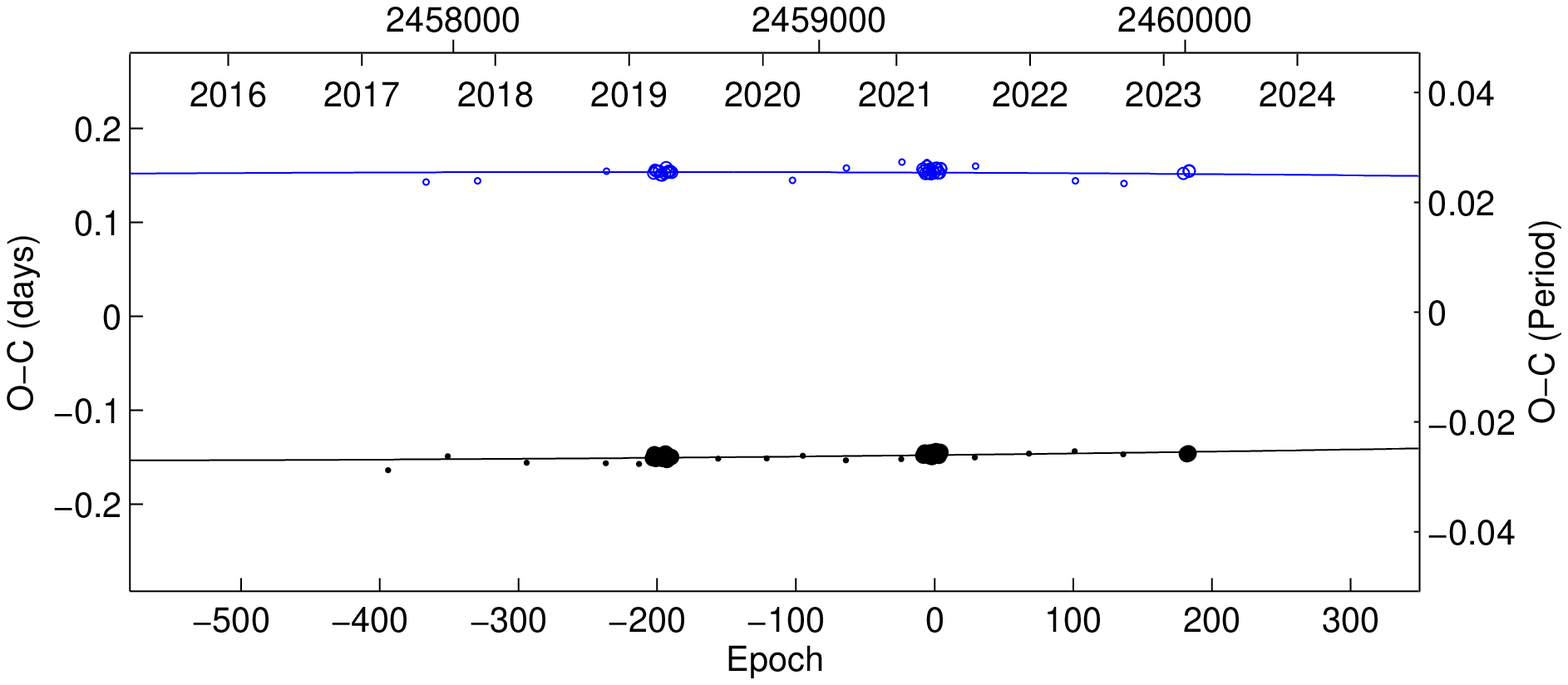}}
 \put(258,120){\includegraphics[width=0.48\textwidth]{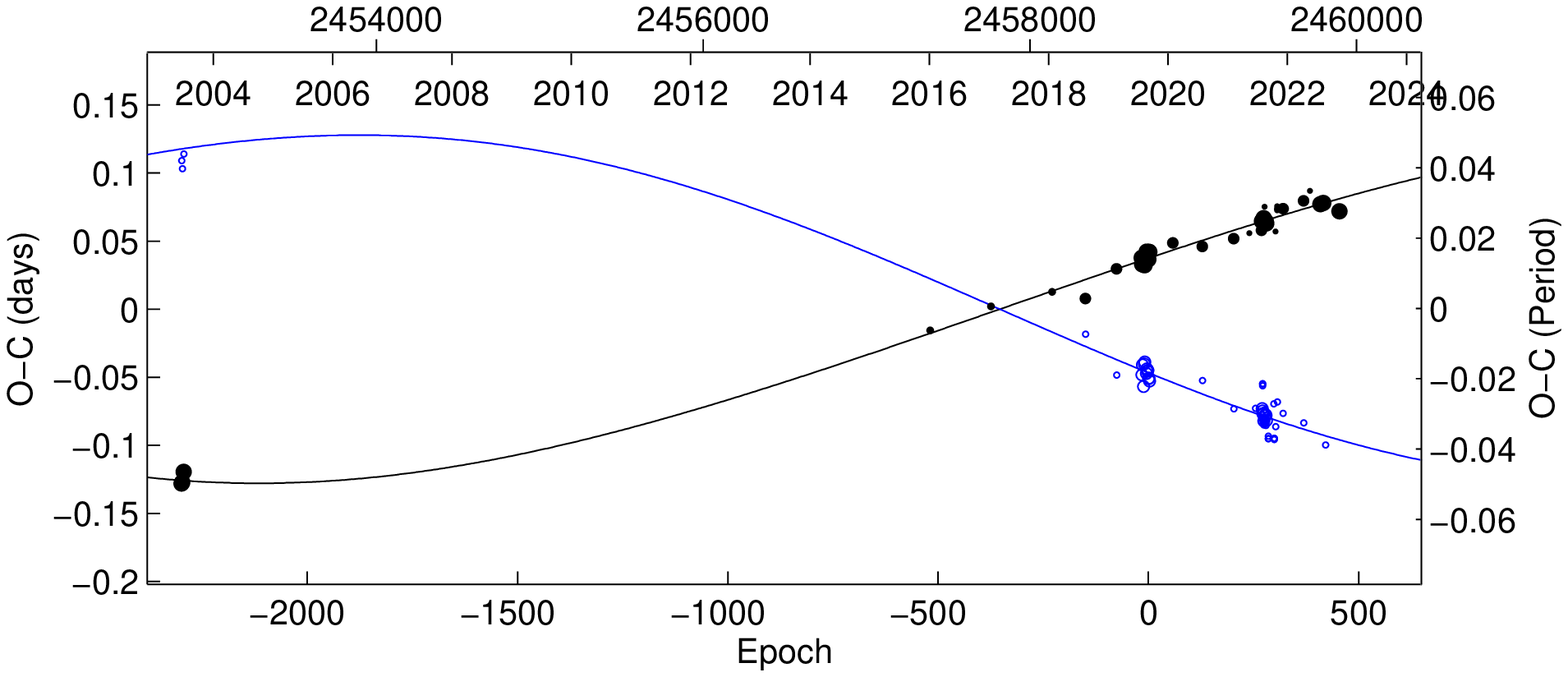}}
 \put(0,0){\includegraphics[width=0.48\textwidth]{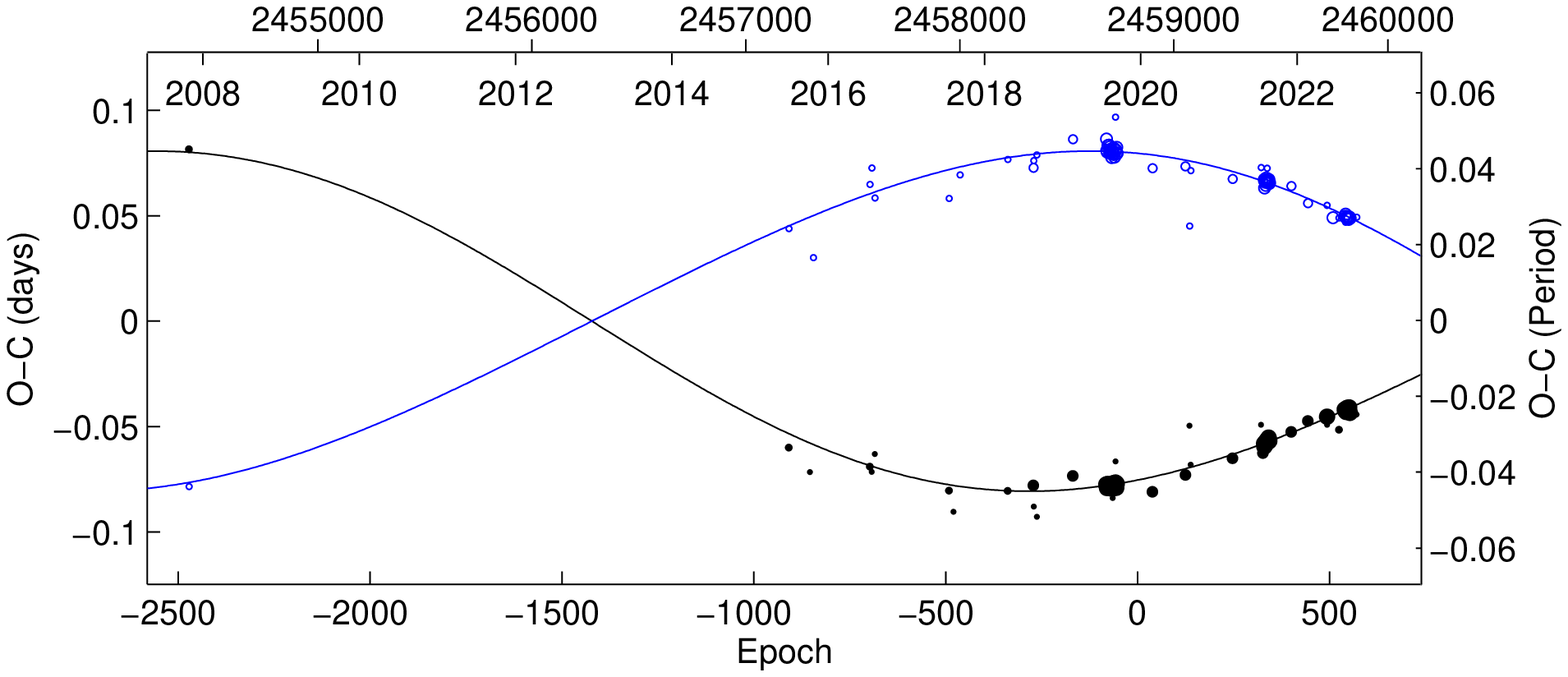}}
 \put(258,0){\includegraphics[width=0.48\textwidth]{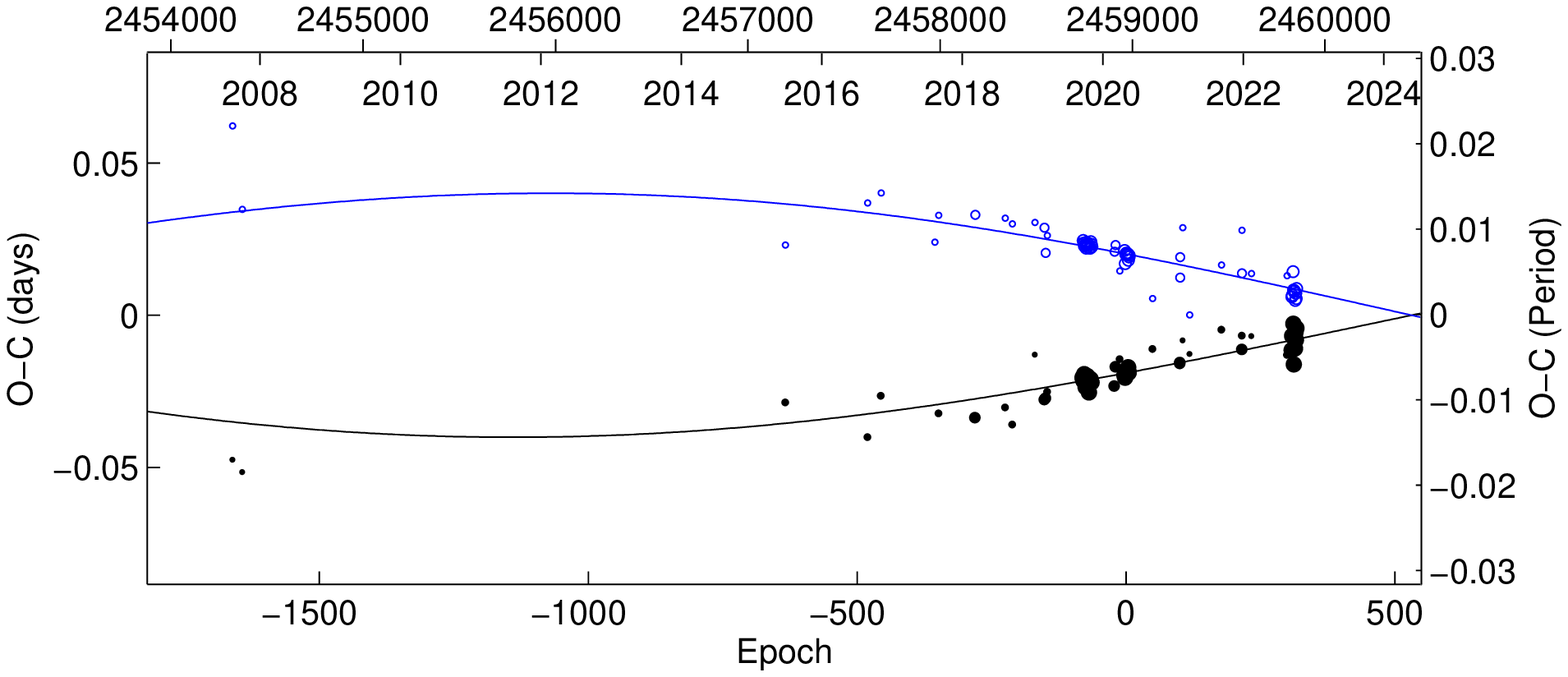}}
   \put(40,141){\small ASASSN-V J102911.57-522413.6: pair B}
   \put(350,141){\small V2894Cyg: pair A}
   \put(40,22){\small NSVS 5725040: pair A}
   \put(325,22){\small ZTF J220518.78+592642.1: pair A}
  \end{picture}
  \caption{Apsidal motion fits of four eccentric orbits: ASASSN-V J102911.57-522413.6: pair B, V2894Cyg: pair A, NSVS 5725040: pair A, and ZTF J220518.78+592642.1: pair A. For all the plots, these diagrams were obtained after subtraction of the ETVs of individual pairs only showing the long-term apsidal motions. Black dots represent the primary eclipses, blue ones are secondaries. }
  \label{Fig_AM}
 \end{figure*}

\section{Discussion and conclusions}  \label{discussion}

We performed the first detailed analysis of seven new multiple systems that have been proven to be
bound quadruples of 2+2 architecture. We were able to detect the period variations of both pairs
of these systems thanks to the collection of photometric observations spanning back several
decades. Despite the fact that their outer mutual periods are sometimes too long, meaning that our
data are still insufficient to fully describe the orbit, our analysis definitively shows  the ETV
variations of both A and B pairs behaving in the opposite manner.

Besides the ETVs of both pairs, we also detected the significant apsidal motions in several
binaries as a byproduct. All of these fits are shown in Figure \ref{Fig_AM}, where we plotted only
the apsidal motion fits following the subtraction of the mutual movement of the individual pairs
around their barycenters.

\begin{figure}
 \centering
 \includegraphics[width=0.45\textwidth]{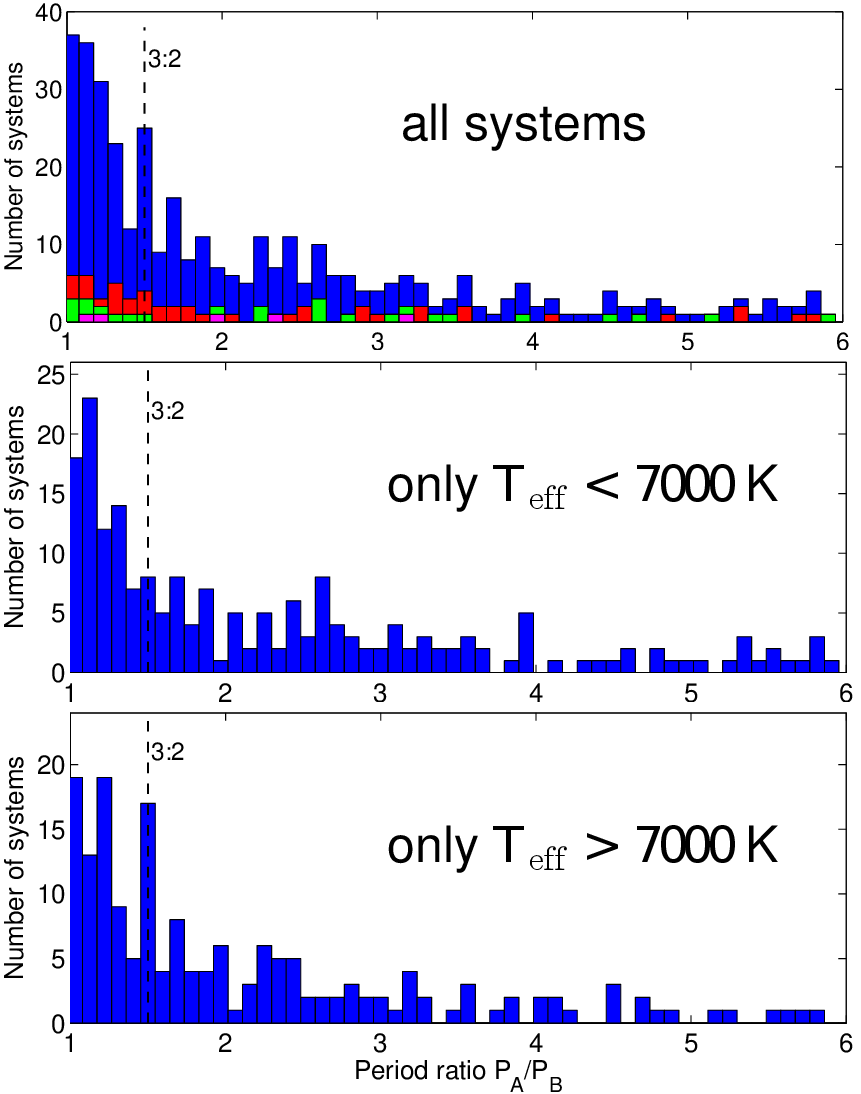}
 \caption{Statistics of the period ratios of both inner pairs. Upper plot: Blue columns showing all 2+2 quadruples, green data showing the other non-eclipsing 2+2 systems from the MSC catalogue, red data are the unconfirmed systems detected but still not published yet, and, finally, the magenta data are our new seven systems presented in the current paper. Middle plot: Only the systems with temperatures below 7000~K. Bottom plot: Only the early-type systems with $T_{eff}>7000$K.}
 \label{Fig_statistics}
\end{figure}

At present, the total number of  doubly eclipsing systems showing definitely two sets of eclipses
is more than 350 (but still only a small fraction of them have been proven to be real bound 2+2
quadruples). As in our previous studies, we plotted the period ratios of all these systems. We
completed the known doubly eclipsing systems with other 2+2 quadruples from the literature and
from the MSC catalog \citep{2018ApJS..235....6T}. This can be seen at the upper part of the Figure
\ref{Fig_statistics}, where the new systems from the present study are shown alongside other
candidates (with two eclipsing periods) that are still unpublished (awaiting publication in the
near future). The number of systems in our sample has reached 450 in total. However, unlike our
previous studies, we also divided the set into two subgroups. At first, the systems of earlier
spectral type, having mostly  radiative atmospheres, with $T_{eff}>7000$K, and/or having a Gaia
photometric index of $(B_p-R_p)<0.45$. Also, on the contrary, the systems having convective
atmospheres of later spectral types, with $T_{eff}<7000$K. These two subplots are also shown in
the lower plots. Surprisingly, the suspicious peak near the 3:2 mean motion resonance is
preferably seen only in the one showing hotter stars with $T_{eff}>7000$K. The question of whether
this indicates some deeper physical reason or it is just a coincidence resulting from the small
number statistics remains open. Finally, we recommend that new systems should be added to extend
the statistics and sample in both these groups.

\begin{acknowledgements}
 We do thank the ASAS, SuperWASP, ZTF, ASAS-SN, and TESS teams for making all of the observations easily public available.
 The research of P.Z. was supported by the project {\sc Cooperatio - Physics} of Charles University in Prague.
 We are also grateful to the ESO team at the La Silla Observatory for their help in maintaining and operating the Danish telescope.
  This work has made use of data from the European Space Agency (ESA) mission {\it Gaia} (\url{https://www.cosmos.esa.int/gaia}), processed by the {\it Gaia}
Data Processing and Analysis Consortium (DPAC,
\url{https://www.cosmos.esa.int/web/gaia/dpac/consortium}). Funding for the DPAC has been provided
by national institutions, in particular the institutions participating in the {\it Gaia}
Multilateral Agreement. We would also like to thank the Pierre Auger Collaboration for the use of
its facilities. The operation of the robotic telescope FRAM is supported by the grant of the
Ministry of Education of the Czech Republic LM2023032. The data calibration and analysis related
to the FRAM telescope is supported by the Ministry of Education of the Czech Republic MSMT-CR
LTT18004, MSMT/EU funds CZ.02.1.01/0.0/0.0/16$\_$013/0001402 and
CZ.02.1.01/0.0/0.0/18$\_$046/0016010. This work is supported by MEYS (Czech Republic) under the
projects MEYS LM2023047, LTT17006 and EU/MEYS CZ.02.1.01/0.0/0.0/16$\_$013/0001403 and
CZ.02.1.01/0.0/0.0/18$\_$046/0016007.
 The research of P.Z., J.K., and J.M. was also supported by the project {\sc Cooperatio - Physics} of Charles University in Prague.
 The observations by Z.H. in Velt\v{e}\v{z}e were obtained with a CCD camera kindly borrowed by the Variable Star and Exoplanet Section of the Czech Astronomical Society.
 This research made use of Lightkurve, a Python package for TESS data analysis \citep{2018ascl.soft12013L}.
 This research has made use of the SIMBAD and VIZIER databases, operated at CDS, Strasbourg, France and of NASA Astrophysics Data System Bibliographic Services.
\end{acknowledgements}

\end{document}